\documentclass[12pt,a4paper,final]{iopart}

\usepackage{graphicx}
\usepackage{sidecap}
\usepackage{datetime}
\usepackage{color}
\usepackage{float}
\usepackage{setspace}
\usepackage{booktabs}
\usepackage{textcomp}
\usepackage{braket}
\expandafter\let\csname equation*\endcsname\relax
\expandafter\let\csname endequation*\endcsname\relax
\usepackage{amsmath}
\usepackage{amssymb}
\usepackage{xr}
\usepackage{iopams}  
\usepackage[breaklinks=true,colorlinks=true,linkcolor=blue,urlcolor=blue,citecolor=blue]{hyperref}

\newcommand{\pdd}[2]{\frac{\partial{#1}}{\partial{#2}}}
\newcommand{\vC}{C}
\newcommand{\gC}{\mathcal{\vC}}

\newcommand{\vK}{K}
\newcommand{\gK}{\mathcal{\vK}}
\newcommand{\vE}{E}

\newcommand{\aE}{e}

\newcommand{\vJ}{J}
\newcommand{\gJ}{\mathcal{\vJ}}
\newcommand{\vL}{D}

\newcommand{\aL}{d}
\newcommand{\aS}{s}

\newcommand{\vM}{M}
\newcommand{\gM}{\mathcal{\vM}}
\newcommand{\vP}{S}
\newcommand{\aP}{\ell}

\usepackage{tikz}
\usetikzlibrary{trees}
\usetikzlibrary{positioning}
\usetikzlibrary{decorations.pathmorphing}
\usetikzlibrary{decorations.markings}
\usetikzlibrary{shapes.geometric}
\usetikzlibrary{snakes}
\usetikzlibrary{arrows}
\usetikzlibrary{calc}

\tikzset{
    full edge/.style={draw=black, solid},
	double edge/.style={draw=black, bend right, solid, double},
    dash edge/.style={draw=black,dashed},
    dot edge/.style={draw=black,dotted},
	snake edge/.style={draw=black,snake=snake, segment amplitude=1pt,segment length=3pt},
    dressed edge/.style={draw=black,solid, postaction={decorate},
        decoration={markings,mark=at position .55 with {\draw[fill=white] circle (2pt);} }},
    dressed double  edge/.style={draw=black,solid, postaction={decorate}, double,
        decoration={markings,mark=at position .55 with {\draw[fill=white] circle (2pt);} }},
    redressed edge/.style={draw=black,solid, postaction={decorate},
        decoration={markings,mark=at position .55 with {\draw[fill=black] circle (2pt);} }},
	redressed double edge/.style={draw=black,solid, postaction={decorate}, double,
        decoration={markings,mark=at position .55 with {\draw[fill=black] circle (2pt);} }},
}
\usetikzlibrary{shapes.geometric}

\tikzset{
    ncbar angle/.initial=90,
    ncbar/.style={
        to path=(\tikztostart)
        -- ($(\tikztostart)!#1!\pgfkeysvalueof{/tikz/ncbar angle}:(\tikztotarget)$)
        -- ($(\tikztotarget)!($(\tikztostart)!#1!\pgfkeysvalueof{/tikz/ncbar angle}:(\tikztotarget)$)!\pgfkeysvalueof{/tikz/ncbar angle}:(\tikztostart)$)
        -- (\tikztotarget)
    },
    ncbar/.default=0.5cm,
}

\tikzset{square left brace/.style={ncbar=0.5cm}}
\tikzset{square right brace/.style={ncbar=-0.5cm}}

\tikzset{round left paren/.style={ncbar=0.5cm,out=120,in=-120}}
\tikzset{round right paren/.style={ncbar=0.5cm,out=60,in=-60}}

\begin{document}

\title{Leaf-to-leaf distances in Catalan trees}
%\title[A correspondence between leaf-to-leaf path length and leaf depth in ordered Catalan tree graphs}

\author{A.M.\ Goldsborough$^{1,2,\ast}$, J.M.\ Fellows$^{2}$, S.A.\ Rautu$^{2,3}$, M.\ Bates$^{2}$, G.\ Rowlands$^{2}$ and R.A.\ R\"{o}mer$^{2,+}$}
\address{$^{1}$Max Planck Institute for Quantum Optics, Hans-Kopfermannstr. 1, D-85748 Garching, Germany}
\address{$^{2}$Department of Physics and Centre for Scientific Computing, University of Warwick, Coventry, CV4 7AL, UK}
\address{$^{3}$Simons Centre for the Study of Living Machines, National Centre for Biological Sciences, Bangalore, Karnataka, India}
%\address{$^{3}$School of Physics, HH Wills Physics Laboratory, University of Bristol, Tyndall Avenue, BS8 1TL, UK}
\ead{$^{\ast}$Andrew.Goldsborough@mpq.mpg.de}
\ead{$^{+}$R.Roemer@warwick.ac.uk}

\begin{abstract}
We study the average leaf-to-leaf path lengths on \emph{ordered} Catalan tree graphs with $n$ nodes and show that these are equivalent to the average length of paths starting from the root node. 
We give an explicit analytic formula for the average leaf-to-leaf path length as a function of separation of the leaves and study its asymptotic properties. 
At the heart of our method is a strategy based on an abstract graph representation of generating functions.
\end{abstract}

\vspace{2pc}
%\maketitle

%%%%%%%%%%%%%%%%%%%%%%%%%%%%%%%%%%%%%%%%%%%%%%%%%%%%%%%%%%%%%%%%%%%
\section{Introduction}
%%%%%%%%%%%%%%%%%%%%%%%%%%%%%%%%%%%%%%%%%%%%%%%%%%%%%%%%%%%%%%%%%%%

The aim of our work is to give an analytic expression for the average path length between leaves in full binary trees.
Similar such tree structures have been used as an analogue for connections in the Hilbert space of wavefunctions in strongly disordered Heisenberg chains \cite{GolR14}, where the path lengths are believed to be related to correlation functions \cite{EveV11}. 
More generally, it has been recently argued that certain RG approaches to the Hilbert space of critical many-body interacting system in $d$ dimensions share many of their geometric properties with $d+1$ dimensional AdS \cite{Swi12,Mol13}. 
This connection is a manifestation of the so-called AdS/CFT correspondence \cite{Mal99} and its possible applications in condensed matter physics \cite{Mcg10}.
In these quantum systems, the leaves are ordered according to their physical position, for example the location of magnetic ions in a quantum wire. 
This ordering imposes a new restriction on the tree itself and the lengths which become important are leaf-to-leaf path lengths across the \emph{ordered} tree. 
Such an ordering is usually not necessary when using traditional measures of path lengths in graph theory and computer science approaches concerned with, e.g., search trees \cite{Ros12}.
%We emphasise that the path lengths therefore correspond to quite different measures than those studied in the traditional graph theory and computer science approaches usually concerned with tree graphs \cite{Ros12}.

The construction of this tree of connections in Hilbert space is based on details of so-called tensor network methods \cite{Oru14,Oru14_2}, most famous of which is the variational matrix-product state (MPS) method that underlies modern density matrix renormalisation group (DMRG) algorithms \cite{Whi92,OstR95,Sch11}. 
These provide elegant and powerful tools for the simulation of quantum many-body systems in low dimensions.
In such loop-free tensor networks, correlations scale as $e^{-\alpha \aP(x_{1},x_{2})}$, where $\aP(x_{1},x_{2})$ is the number of tensors that connect leaf $x_{1}$ to leaf $x_{2}$ \cite{EveV11}.
Typical MPS methods for gapped systems give $\aP \approx |x_{2}-x_{1}|$ and correlations scale exponentially. 
On the other hand a path length $\aP \approx \text{log} |x_{2}-x_{1}|$ as found in the multi-scale entanglement renormalisation ansatz (MERA) \cite{Vid07,Vid08_2,EveV09}, leads to a power law decay $\sim e^{-\alpha \text{log}|x_{2} - x_{1}|} \sim |x_{2} - x_{1}|^{-\alpha}$.

%%%%%%%%%%%%%%%%%%%%%%%%%%%%%%%%%%%%%%%%
\begin{figure}[ht]
    \centering
    \includegraphics[width=0.7\columnwidth]{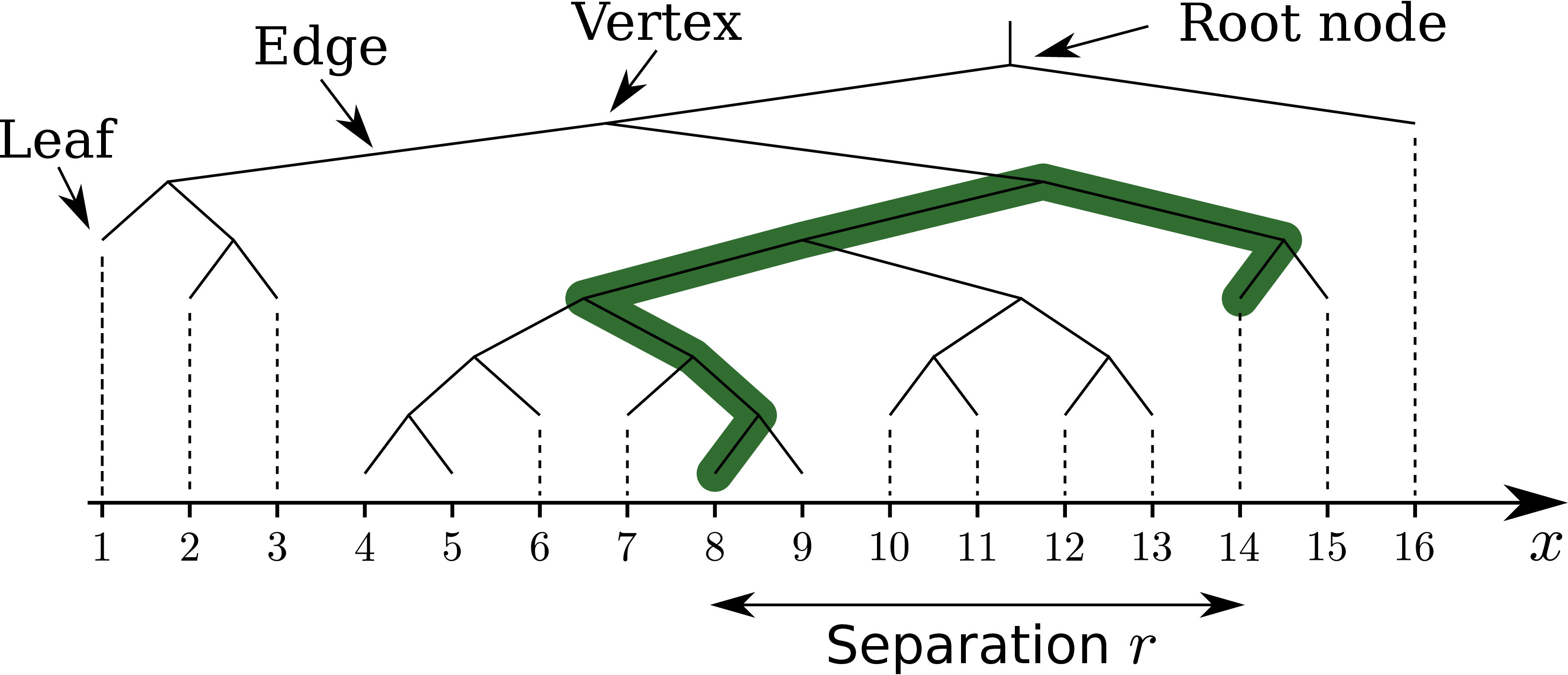}
    \caption{A Catalan tree graph with various definitions as defined in the main text.
    %Circles ($\bullet$, $\circ$) denote vertices while lines indicate edges between the vertices of different depth. 
    The tree as shown has $n=15$ and $16$ leaves. The indicated separation is $r=6$ while the associated leaf-to-leaf path length equals $\ell=6$ as indicated by the thick line. The thick green line denotes the path.
    \label{fig-schematic}}
\end{figure}
%%%%%%%%%%%%%%%%%%%%%%%%%%%%%%%%%%%%%%%%
Different quantum systems give rise to different tree structures, and these in turn lead to different distance behaviours for the path lengths. 
Hence it is interesting to ask more generally about the path lengths of different tree graphs. 
In a previous work \cite{GolRR15}, three of us studied the case of full \emph{and} complete $m$-ary tree graphs using a recursive approach. 
We found explicit expressions in terms of Hurwitz-Lerch transcendants. 
For incomplete random binary trees, we presented first numerical results. 
In the present manuscript, we extend our results to the set of all full binary trees (the so-called \emph{Catalan tree graphs}), i.e.\ we drop the requirement of completeness.
An example of path lengths in such Catalan trees is given in Fig.\ \ref{fig-schematic}.
The absence of paths that would be present in a complete binary tree seems similar to the more random structure seen in random binary trees \cite{GolRR15}. 
We might hence expect that Catalan trees should in many ways be similar to the tree structure found previously \cite{GolR14}.

We consider a tree graph where every internal vertex has exactly two children as shown in Fig.\ \ref{fig-schematic}.
%%%%%%%%%%%%%%%%%%%%%%%%%%%%%%%%%%%%%%%%
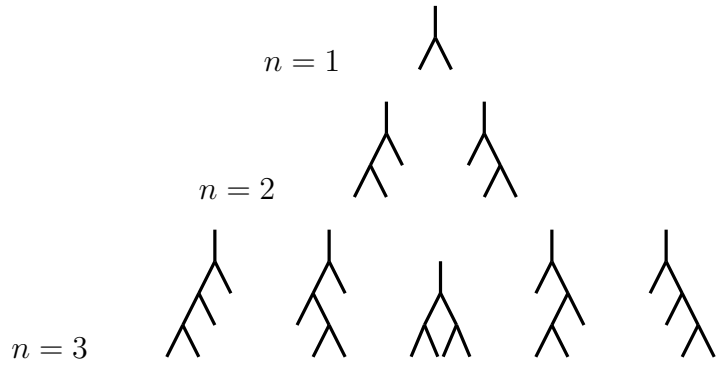
\begin{figure}[ht]
    \centering
    $n = 1$ \qquad
    \begin{tikzpicture} [very thick,scale=0.5]
    	\tikzstyle{level 1}=[level distance=24pt]
    	\tikzstyle{level 2}=[level distance=24pt, sibling distance=24pt]
    	\coordinate
    		child{ edge from parent [full edge]
    			child{ edge from parent [full edge] }
    			child{ edge from parent [full edge] }
    		} ;
    \end{tikzpicture}
    \vspace{10pt}\\
    $n = 2$ \qquad
    \begin{tikzpicture} [very thick,scale=0.5]
    	\tikzstyle{level 1}=[level distance=24pt]
    	\tikzstyle{level 2}=[level distance=24pt, sibling distance=24pt]
    	\tikzstyle{level 3}=[level distance=24pt, sibling distance=24pt]
    	\coordinate
    		child{ edge from parent [full edge]
    			child{ edge from parent [full edge]
    				child{ edge from parent [full edge] }
    				child{ edge from parent [full edge] }			
    			}
    			child{ edge from parent [full edge] }
    		} ;
    \end{tikzpicture}
    ~~~~
    \begin{tikzpicture} [very thick,scale=0.5]
    	\tikzstyle{level 1}=[level distance=24pt]
    	\tikzstyle{level 2}=[level distance=24pt, sibling distance=24pt]
    	\tikzstyle{level 3}=[level distance=24pt, sibling distance=24pt]
    	\coordinate
    		child{ edge from parent [full edge]
    			child{ edge from parent [full edge] }
    			child{ edge from parent [full edge]
    				child{ edge from parent [full edge] }
    				child{ edge from parent [full edge] }			
    			}
    		} ;
    \end{tikzpicture}
    \vspace{10pt}\\
    $n = 3$ \qquad
    \begin{tikzpicture} [very thick,scale=0.5]
    	\tikzstyle{level 1}=[level distance=24pt]
    	\tikzstyle{level 2}=[level distance=24pt, sibling distance=24pt]
    	\tikzstyle{level 3}=[level distance=24pt, sibling distance=24pt]
    	\tikzstyle{level 4}=[level distance=24pt, sibling distance=24pt]
    	\coordinate
    		child{ edge from parent [full edge]
    			child{ edge from parent [full edge]
    				child{ edge from parent [full edge] 	
    				    child{ edge from parent [full edge] }
    				    child{ edge from parent [full edge] }			
    			        }
    				child{ edge from parent [full edge] }
    		        }
    		    child{ edge from parent [full edge] }
                };
    \end{tikzpicture}
    ~~~~
    \begin{tikzpicture} [very thick,scale=0.5]
    	\tikzstyle{level 1}=[level distance=24pt]
    	\tikzstyle{level 2}=[level distance=24pt, sibling distance=24pt]
    	\tikzstyle{level 3}=[level distance=24pt, sibling distance=24pt]
    	\tikzstyle{level 4}=[level distance=24pt, sibling distance=24pt]
    	\coordinate
    		child{ edge from parent [full edge]
    			child{ edge from parent [full edge]
    				child{ edge from parent [full edge] } 	
    				child{ edge from parent [full edge] 
    				    child{ edge from parent [full edge] }
    				    child{ edge from parent [full edge] }			
                        }
    		        }
    		    child{ edge from parent [full edge] }
                };
    \end{tikzpicture}
    ~~~~
    \begin{tikzpicture} [very thick,scale=0.5]
    	\tikzstyle{level 1}=[level distance=24pt]
    	\tikzstyle{level 2}=[level distance=24pt, sibling distance=24pt]
    	\tikzstyle{level 3}=[level distance=24pt, sibling distance=20pt]
    	\coordinate
    		child{ edge from parent [full edge]
    			child{ edge from parent [full edge]
    				child{ edge from parent [full edge] }
    				child{ edge from parent [full edge] }			
    			}
    			child{ edge from parent [full edge]
    				child{ edge from parent [full edge] }
    				child{ edge from parent [full edge] }			
    			}
    		} ;
    \end{tikzpicture}
    ~~~~
    \begin{tikzpicture} [very thick,scale=0.5]
    	\tikzstyle{level 1}=[level distance=24pt]
    	\tikzstyle{level 2}=[level distance=24pt, sibling distance=24pt]
    	\tikzstyle{level 3}=[level distance=24pt, sibling distance=24pt]
    	\tikzstyle{level 4}=[level distance=24pt, sibling distance=24pt]
    	\coordinate
    		child{ edge from parent [full edge]
    		    child{ edge from parent [full edge] }
    			child{ edge from parent [full edge]
    				child{ edge from parent [full edge] 	
    				    child{ edge from parent [full edge] }
    				    child{ edge from parent [full edge] }			
    			    }
    				child{ edge from parent [full edge] }
    		    }
            };
    \end{tikzpicture}
    ~~~~
    \begin{tikzpicture} [very thick,scale=0.5]
    	\tikzstyle{level 1}=[level distance=24pt]
    	\tikzstyle{level 2}=[level distance=24pt, sibling distance=24pt]
    	\tikzstyle{level 3}=[level distance=24pt, sibling distance=24pt]
    	\tikzstyle{level 4}=[level distance=24pt, sibling distance=24pt]
    	\coordinate
    		child{ edge from parent [full edge]
    		    child{ edge from parent [full edge] }
    			child{ edge from parent [full edge]
    				child{ edge from parent [full edge] }
    				child{ edge from parent [full edge] 	
    				    child{ edge from parent [full edge] }
    				    child{ edge from parent [full edge] }			
    			    }
    		    }
            };
    \end{tikzpicture}
    \caption{The set of all full binary trees with $n=1$, $2$ and $3$ vertices. 
    \label{fig:n_123_all}}
\end{figure}
%%%%%%%%%%%%%%%%%%%%%%%%%%%%%%%%%%%%%%%%
Such a tree is usually called a \emph{full binary tree}.  
Let us define $n$ as the number of \emph{internal} vertices in the graph. 
We shall refer to such internal vertices as \emph{nodes} or simply \emph{vertices}. 
The \emph{root} is the top vertex of the tree, which is unique as it is not child of any vertex.
We shall call a terminal vertex with no children a \emph{leaf}.

The number of full binary trees with $n$ vertices is given by the {Catalan number} $\vC_{n}$, giving rise to the set being known as the \emph{Catalan tree graphs} \cite{SedF13}.  
It is straightforward to show that with $L$ denoting the number of leaves, then $L=n+1$.
For the examples given in Fig.\ \ref{fig:n_123_all}, we see that for $n=1$, $2$, and $3$, we have $\vC_1= 1$, $\vC_2=2$ and $\vC_3=5$ unique trees with $2$, $3$
and $4$ leaves in each such tree.

%%%%%%%%%%%%%%%%%%%%%%%%%%%%%%%%%%%%%%%%%%%%%%%%%%%%%%%%%%%%%%%%%%%
\section{Catalan numbers and their generating functions in ordered Catalan trees}
%%%%%%%%%%%%%%%%%%%%%%%%%%%%%%%%%%%%%%%%%%%%%%%%%%%%%%%%%%%%%%%%%%%

%%%%%%%%%%%%%%%%%%%%%%%%%%%%%%%%%%%%%%%%%%%%%%%
\subsection{Basic properties of Catalan numbers and the diagrammatic approach}
%%%%%%%%%%%%%%%%%%%%%%%%%%%%%%%%%%%%%%%%%%%%%%%
The theory of Catalan numbers and their applications is well-developed \cite{Kos09}, so we only briefly cite some of the results we shall be needing
in the following. 
Catalan numbers can be computed explicitly using well-known relations such as, e.g.
\begin{equation}
    \vC_n = \frac{1}{n+1}\binom{2n}{n} = \frac{(2n)!}{(n+1)!n!}
           = \frac{4n-2}{n+1} \vC_{n-1} = \sum_{\alpha=0}^{n-1} \vC_{\alpha} \vC_{n-\alpha-1} ~,
    \label{eq-Catalan_recursion}
\end{equation}
where the last equation in \eqref{eq-Csum} is a form of Segner's recurrence relation \cite{Kos09}.
% In the following, we shall make extensive use of generating functions \cite{Kos09,Wil94}.
% Generating functions offer a convenient way to manipulate relations between the set of Catalan numbers. 
% More generally, they serve as a very useful formal device for manipulating a series and for determining the properties of unknown series \cite{Wil94}.
% Given a general (infinite) series $a_0,a_1,a_2,a_3\dots$, it is always possible to formally define a function
% \begin{equation}
% \label{eq-defgenfun}
% a(x) = a_0 + a_1 x + a_2 x^2 + a_3 x^3 + \dots = \sum_{n=0}^{\infty} a_n x^n ~,
% \end{equation}
% which we call the \emph{generating function} of the sequence $a_n$.
% We can also invert the above definition and instead say that $a_n$ is the coefficient of $x^n$ in the Taylor series representation of $a(x)$ about zero. 
% Following Ref.\ \cite{Wil94}, we will employ the notation $a_n = [x^n] \{ a(x) \}$.
Alternatively, we can define the $\vC_n$'s via generating function as
\begin{equation}
    \gC(x) = \sum_{n=0}^{\infty} \vC_{n} x^{n} ~,
\label{eq-Csum}
\end{equation}
and, using established notation \cite{Wil94}, we write the inverse relation as 
%\begin{equation}
 $\vC_{n} = [x^n]\{ \gC(x) \}$.
%\label{eq-CnfromgC}
%\end{equation}
% Squaring \eqref{eq-Csum} gives,
% \begin{align}
% \gC^{2}(x) &= \vC_{0}^{2} + (\vC_{0}\vC_{1} + \vC_{1}\vC_{0})x + \dots + (\vC_{0}\vC_{n} + \vC_{1}\vC_{n-1} + \dots + \vC_{n}\vC_{0})x^{n} + \dots \nonumber \\
%   &= \vC_{1} + \vC_{2}x + \vC_{3}x^{2} + \dots + \vC_{n+1}x^{n} + \dots \nonumber \\
%   &= \frac{\gC(x)-\vC_{0}}{x} ,
%   \label{eq-Cderiv}
% \end{align}
% and we have used the final part of Eq.\ \eqref{eq-Catalan_recursion} in each term.
% Multiplication by $x$ results in
It is straightforward to show
\begin{equation}
    \gC(x) = 1+ x\gC^{2}(x) ~,
\label{eq-Cquad}
\end{equation}
which is the generating function version of Segner's relation and a useful form for subsequent calculations.
More algebraic details of generating functions are provided in the appendix \ref{sec:generating_functions}.
Next, we introduce a diagrammatic notation \cite{Kir83,Wil94} for the generating function in which a vertex \emph{represents} $x$ so that we can directly add together diagrams. 
% This will provide an intuition for the development of the generating functions as we can manipulate the diagrams and replace them with appropriate terms at the end.
We begin by setting up a diagrammatic dictionary such that
\begin{equation}
    \begin{tikzpicture} [thick,scale=0.5]
    	\tikzstyle{level 1}=[level distance=24pt]
    	\coordinate
    		child{ edge from parent [full edge]
    		} ;
    \end{tikzpicture} \quad \triangleq 1 \quad , \qquad
    \begin{tikzpicture} [thick,scale=0.5]
    	\tikzstyle{level 1}=[level distance=24pt]
    	\tikzstyle{level 2}=[level distance=24pt, sibling distance=24pt]
    	\coordinate
    		child{ edge from parent [full edge]
    			child{ edge from parent [full edge] }
    			child{ edge from parent [full edge] }
    		} ;
    \end{tikzpicture} \quad \triangleq x \quad , \qquad
    \begin{tikzpicture} [thick,scale=0.5]	
    	\tikzstyle{level 1}=[level distance=24pt]
    	\coordinate
    		child{ edge from parent [full edge] circle(3pt) } ;	
    \end{tikzpicture} \quad \triangleq {\gC}(x) ~.
\end{equation}
This dictionary allows us to construct a diagrammatic equation for $\gC(x)$, i.e.\
\begin{align}
    \begin{tikzpicture} [thick,scale=0.5]	
    	\tikzstyle{level 1}=[level distance=24pt]
    	\coordinate
    			child{ edge from parent [full edge]  circle(3pt) } ;	
    \end{tikzpicture}
    ~~&=~~
    \begin{tikzpicture} [thick,scale=0.5]
    	\tikzstyle{level 1}=[level distance=24pt]
    	\coordinate
    		child{ edge from parent [full edge]
    		} ;
    \end{tikzpicture}
    ~~+~~
    \begin{tikzpicture} [thick,scale=0.5]
    	\tikzstyle{level 1}=[level distance=24pt]
    	\tikzstyle{level 2}=[level distance=24pt, sibling distance=24pt]
    	\coordinate
    		child{ edge from parent [full edge]
    			child{ edge from parent [full edge] }
    			child{ edge from parent [full edge] }
    		} ;
    \end{tikzpicture}
    ~~+~~
    \begin{tikzpicture} [thick,scale=0.5]
    	\tikzstyle{level 1}=[level distance=24pt]
    	\tikzstyle{level 2}=[level distance=24pt, sibling distance=24pt]
    	\tikzstyle{level 3}=[level distance=24pt, sibling distance=24pt]
    	\coordinate
    		child{ edge from parent [full edge]
    			child{ edge from parent [full edge]
    				child{ edge from parent [full edge] }
    				child{ edge from parent [full edge] }			
    			}
    			child{ edge from parent [full edge] }
    		} ;
    \end{tikzpicture}
    ~~+~~
    \begin{tikzpicture} [thick,scale=0.5]
    	\tikzstyle{level 1}=[level distance=24pt]
    	\tikzstyle{level 2}=[level distance=24pt, sibling distance=24pt]
    	\tikzstyle{level 3}=[level distance=24pt, sibling distance=24pt]
    	\coordinate
    		child{ edge from parent [full edge]
    			child{ edge from parent [full edge] }
    			child{ edge from parent [full edge]
    				child{ edge from parent [full edge] }
    				child{ edge from parent [full edge] }			
    			}
    		} ;
    \end{tikzpicture}
    ~~+~~
    \begin{tikzpicture} [thick,scale=0.5]
    	\tikzstyle{level 1}=[level distance=24pt]
    	\tikzstyle{level 2}=[level distance=24pt, sibling distance=24pt]
    	\tikzstyle{level 3}=[level distance=24pt, sibling distance=20pt]
    	\coordinate
    		child{ edge from parent [full edge]
    			child{ edge from parent [full edge]
    				child{ edge from parent [full edge] }
    				child{ edge from parent [full edge] }			
    			}
    			child{ edge from parent [full edge]
    				child{ edge from parent [full edge] }
    				child{ edge from parent [full edge] }			
    			}
    		} ;
    \end{tikzpicture}
    ~~+\dots \nonumber\\
    &= 1 + x + x^2 + x^2+ x^3 + \ldots %\nonumber \\
    %& 
    = 1 + x + 2x^2 + 5 x^3 + \ldots ~,
\end{align}
which contains all of the trees shown in Fig.\ \ref{fig:n_123_all}.
We also notice that $\gC(x)$ can be generated recursively according to the formula
\begin{align}
    \begin{tikzpicture} [thick,scale=0.75]	
    	\tikzstyle{level 1}=[level distance=24pt]
    	\coordinate
    		child{ edge from parent [full edge]  circle(3pt) } ;		
    \end{tikzpicture}
    ~~&=~~
    \begin{tikzpicture} [thick,scale=0.75]
    	\tikzstyle{level 1}=[level distance=24pt]
    	\coordinate
    		child{ edge from parent [full edge]
    		} ;
    \end{tikzpicture}
    ~~+~~
    \begin{tikzpicture} [thick,scale=0.75]
    	\tikzstyle{level 1}=[level distance=24pt]
    	\tikzstyle{level 2}=[level distance=24pt, sibling distance=24pt]
    	\coordinate
    		child{ edge from parent [full edge]
    			child{ edge from parent [full edge]  circle(3pt) }
    			child{ edge from parent [full edge]  circle(3pt)  }
    		} ;
    \end{tikzpicture} ~.
\end{align}
Translating back into algebra, we recover the fundamental relation \eqref{eq-Cquad}.

%%%%%%%%%%%%%%%%%%%%%%%%%%%%%%%%%%%%%%%%%%%%%%%%%%%%%%%%%%%%%%%%%%%
\section{The average depth of leaves in Catalan trees}
%%%%%%%%%%%%%%%%%%%%%%%%%%%%%%%%%%%%%%%%%%%%%%%%%%%%%%%%%%%%%%%%%%%
Before approaching the problem of average leaf-to-leaf path lengths we find it necessary to first determine the average depth of a leaf in a Catalan tree. This problem has been studied previously \cite{Kir83}, we reconsider it here to establish the notation and introduce our approach.
% \footnote{We only discovered Kirchenhoffer's proof after finishing the current work. We also found that obtaining this original paper is very difficult, so providing a separate proof may be valuable to those also in this position.}

%%%%%%%%%%%%%%%%%%%%%%%%%%%%%%%%%%%%%%%%%%%%%%%
\subsection{External leaf depth}
\label{sec:external_paths}
%%%%%%%%%%%%%%%%%%%%%%%%%%%%%%%%%%%%%%%%%%%%%%%
We first consider the simplest case; where the leaf under consideration is the leftmost (or equivalently rightmost).
We approach this by finding how many binary trees with $n$ vertices there are such that the leftmost leaf is $k$ vertices away from the root. 
We will call such a configuration an \emph{external} (because it is on the outer left leaf) \emph{rooted} (because it starts at the root) \emph{path} of length $k$ on a binary tree with $n$ vertices.
We also note that due to symmetry reasons the results should hold as well for the \emph{rightmost} leaf of the tree.
We will denote by $\vK_{jk}$ the number of external rooted paths of length $k$ on a binary tree with $n=j+k$ vertices (so that there are always at least $k$ vertices). 
We also define its generating function as ${\gK}(x,z)=\sum_{jk}\vK_{jk}x^jz^k$.
In order to highlight the path along the leftmost leaves we will define a new diagrammatic dictionary
\begin{equation}
    \quad
    \begin{tikzpicture} [thick,scale=0.95]
    	\tikzstyle{level 1}=[level distance=24pt]
    	\coordinate
    		child{ edge from parent [full edge]
    		} ;
    \end{tikzpicture} \quad = \quad
    \begin{tikzpicture} [thick,scale=0.95]
    	\tikzstyle{level 1}=[level distance=24pt]
    	\coordinate
    		child{ edge from parent [dash edge]
    		} ;
        \end{tikzpicture}  \quad \triangleq 1 \quad , \quad
    \begin{tikzpicture} [thick,scale=0.5]
    	\tikzstyle{level 1}=[level distance=24pt]
    	\tikzstyle{level 2}=[level distance=24pt, sibling distance=24pt]
    	\coordinate
    		child{ edge from parent [dash edge]
    			child{ edge from parent [dash edge] }
    			child{ edge from parent [dash edge] }
    		} ;
        \end{tikzpicture} \triangleq x \quad , \quad
    \begin{tikzpicture} [thick,scale=0.5]
    	\tikzstyle{level 1}=[level distance=24pt]
    	\tikzstyle{level 2}=[level distance=24pt, sibling distance=24pt]
    	\coordinate
    		child{ edge from parent [full edge]
    			child{ edge from parent [full edge] }
    			child{ edge from parent [dash edge] }
    		} ;
    \end{tikzpicture} \triangleq z \quad , \quad 
    \begin{tikzpicture} [thick,scale=0.5]	
    	\tikzstyle{level 1}=[level distance=24pt]
    	\coordinate
    		child{ edge from parent [dash edge]  circle(3pt) } ;	
    \end{tikzpicture} \triangleq {\gC}(x)\quad , \quad
    %& \vspace{-10pt}
    \begin{tikzpicture} [thick,scale=0.75]
    	\tikzstyle{level 1}=[level distance=24pt]
    	\coordinate
    		child{ edge from parent [dressed edge] } ;
    \end{tikzpicture}  \triangleq {\gK}(x,z) ~.
    %& \mbox{ }
\end{equation}
Here the $z$ argument allows us to count the number of vertices on (left) external rooted paths while the $x$ term denotes the remaining ones (to the right).
The diagrammatic expression for ${\gK}(x,z)$ is then
\begin{align}
    \begin{tikzpicture} [thick,scale=0.5]	
    	\tikzstyle{level 1}=[level distance=24pt]
    	\coordinate
    			child{ edge from parent [dressed edge] } ;	
    \end{tikzpicture}
    ~~&=~~
    \begin{tikzpicture} [thick,scale=0.5]
    	\tikzstyle{level 1}=[level distance=24pt]
    	\coordinate
    		child{ edge from parent [full edge]
    		} ;
    \end{tikzpicture}
    ~~+~~
    \begin{tikzpicture} [thick,scale=0.5]
    	\tikzstyle{level 1}=[level distance=24pt]
    	\tikzstyle{level 2}=[level distance=24pt, sibling distance=24pt]
    	\coordinate
    		child{ edge from parent [full edge]
    			child{ edge from parent [full edge] }
    			child{ edge from parent [dash edge] }
    		} ;
    \end{tikzpicture}
    ~~+~~
    \begin{tikzpicture} [thick,scale=0.5]
    	\tikzstyle{level 1}=[level distance=24pt]
    	\tikzstyle{level 2}=[level distance=24pt, sibling distance=24pt]
    	\tikzstyle{level 3}=[level distance=24pt, sibling distance=24pt]
    	\coordinate
    		child{ edge from parent [full edge]
    			child{ edge from parent [full edge]
    				child{ edge from parent [full edge] }
    				child{ edge from parent [dash edge] }			
    			}
    			child{ edge from parent [dash edge] }
    		} ;
    \end{tikzpicture}
    ~~+~~
    \begin{tikzpicture} [thick,scale=0.5]
    	\tikzstyle{level 1}=[level distance=24pt]
    	\tikzstyle{level 2}=[level distance=24pt, sibling distance=24pt]
    	\tikzstyle{level 3}=[level distance=24pt, sibling distance=24pt]
    	\coordinate
    		child{ edge from parent [full edge]
    			child{ edge from parent [full edge] }
    			child{ edge from parent [dash edge]
    				child{ edge from parent [dash edge] }
    				child{ edge from parent [dash edge] }			
    			}
    		} ;
    \end{tikzpicture}
    ~~+~~
    \begin{tikzpicture} [thick,scale=0.5]
    	\tikzstyle{level 1}=[level distance=24pt]
    	\tikzstyle{level 2}=[level distance=24pt, sibling distance=24pt]
    	\tikzstyle{level 3}=[level distance=24pt, sibling distance=20pt]
    	\coordinate
    		child{ edge from parent [full edge]
    			child{ edge from parent [full edge]
    				child{ edge from parent [full edge] }
    				child{ edge from parent [dash edge] }			
    			}
    			child{ edge from parent [dash edge]
    				child{ edge from parent [dash edge] }
    				child{ edge from parent [dash edge] }			
    			}
    		} ;
    \end{tikzpicture}
    ~~+\dots
\end{align}
From the results of the previous section we can sum over all of the $x$ vertices to get to
\begin{align}
    \begin{tikzpicture} [thick,scale=0.5]	
    	\tikzstyle{level 1}=[level distance=24pt]
    	\coordinate
    			child{ edge from parent [dressed edge] } ;	
    \end{tikzpicture}
    ~~&=~~
    \begin{tikzpicture} [thick,scale=0.5]
    	\tikzstyle{level 1}=[level distance=24pt]
    	\coordinate
    		child{ edge from parent [full edge]
    		} ;
    \end{tikzpicture}
    ~~+~~
    \begin{tikzpicture} [thick,scale=0.5]
    	\tikzstyle{level 1}=[level distance=24pt]
    	\tikzstyle{level 2}=[level distance=24pt, sibling distance=24pt]
    	\coordinate
    		child{ edge from parent [full edge]
    			child{ edge from parent [full edge] }
    			child{ edge from parent [dash edge] circle(3pt) }
    		} ;
    \end{tikzpicture}
    ~~+~~
    \begin{tikzpicture} [thick,scale=0.5]
    	\tikzstyle{level 1}=[level distance=24pt]
    	\tikzstyle{level 2}=[level distance=24pt, sibling distance=24pt]
    	\tikzstyle{level 3}=[level distance=24pt, sibling distance=24pt]
    	\coordinate
    		child{ edge from parent [full edge]
    			child{ edge from parent [full edge]
    				child{ edge from parent [full edge] }
    				child{ edge from parent [dash edge]  circle(3pt) }			
    			}
    			child{ edge from parent [dash edge]  circle(3pt) }
    		} ;
    \end{tikzpicture}
    ~~+\dots ~.
    \label{eq-kdiagram}
\end{align}
At this point we recognise that the diagrammatic expression \eqref{eq-kdiagram} is generated by the recursion relation
\begin{align}
    \begin{tikzpicture} [thick,scale=0.5]	
    	\tikzstyle{level 1}=[level distance=24pt]
    	\coordinate
    			child{ edge from parent [dressed edge] } ;	
    \end{tikzpicture}
    ~~&=~~
    \begin{tikzpicture} [thick,scale=0.5]
    	\tikzstyle{level 1}=[level distance=24pt]
    	\coordinate
    		child{ edge from parent [full edge]
    		} ;
    \end{tikzpicture}
    ~~+~~
    \begin{tikzpicture} [thick,scale=0.5]
    	\tikzstyle{level 1}=[level distance=24pt]
    	\tikzstyle{level 2}=[level distance=24pt, sibling distance=24pt]
    	\coordinate
    		child{ edge from parent [full edge]
    			child{ edge from parent [dressed edge] }
    			child{ edge from parent [dash edge]  circle(3pt) }
    		} ;
    \end{tikzpicture} ~,
\end{align}
which we translate back into algebra according to our diagrammatic dictionary to finally arrive at
\begin{equation}
    {\gK}(x,z) = 1 + z {\gC}(x) {\gK}(x,z) = \frac{1}{1-z {\gC}(x)} ~.
\end{equation}
We want to know the average depth of an external leaf on a binary tree with $n$ vertices, $\aE_n$. 
This can be found by calculating the sum of external leaf depths $\vE_n$ and dividing by the number of possible trees, $\vC_n$.
We can calculate $\vE_n$ by multiplying the number of external rooted paths with length $k$ by $k$ and summing as follows
\begin{align}
    \vE_n &= \sum_{k=1}^n k \vK_{n-k,k} 
    =\sum_{k=1}^n k[x^{n-k}z^k] \{ {\gK}(x,z) \} \nonumber \\
    &= \sum_{k=1}^n [x^{n-k}z^k] \left\{z \pdd{}{z} {\gK}(x,z) \right\} 
    = \sum_{k=1}^n [x^{n-k}z^k] \{ z{\gC}(x){\gK}(x,z)^2 \} \nonumber \\
    &= [x^n] \{x{\gC}(x)^3 \}
    = \vC_{n+1} - \vC_{n} ~,
    \label{eq-vEn}
\end{align}
where we have used equations \eqref{eq-Cquad}, \eqref{eq-sumout}, \eqref{eq-genfunlower}, \eqref{eq-genfuneuler}, and \eqref{eq-kxx}.
The average depth $\aE_n$ is then found to be 
\begin{equation}
\aE_n = \frac{\vE_n}{\vC_n}
= \frac{\vC_{n+1}}{\vC_n}-1 
= \frac{3n}{n+2} ~.
\end{equation}
In the limit $n\rightarrow\infty$ we find that the average depth approaches $\aE_{\infty}=3$.

%%%%%%%%%%%%%%%%%%%%%%%%%%%%%%%%%%%%%%%%%%%%%%%
\subsection{General leaf depths}
%%%%%%%%%%%%%%%%%%%%%%%%%%%%%%%%%%%%%%%%%%%%%%%
Now we generalise the previous result to look at paths which pass through the tree rather than just along the outer edge. 
We ask how many binary trees with $n$ vertices there are such that the $p^\text{th}$ leaf in from the left is $m$ vertices away from the root. 
We will call this configuration a \emph{rooted path of penetration} $p$ on a binary tree with $n$ vertices. 
According to this definition, an external rooted path would be a rooted path with penetration $p=0$.
In order to understand the penetration we need to distinguish between those vertices to the left of the path and those vertices to the right of the path. 
We will also need to distinguish between those vertices on the path that veer to the left and those that veer to the right.

With this in mind we will construct the generating function $\gJ(\xi,x,z,\zeta)$ which enumerates those trees containing, (i) $i$ vertices to the left of the path, (ii) $j$ vertices to the right of the path, (iii) $k$ left-turning vertices within the path, and (iv) $l$ right-turning vertices within the path.
We will call the series generated $\vJ_{ijkl}=[\xi^i x^j z^k \zeta^ l]\{ {\gJ}(\xi,x,z,\zeta) \}$.
Now the diagrammatic method becomes particularly useful. 
We define the diagrammatic dictionary
\begin{align}
    \begin{tikzpicture} [thick,scale=0.95]
    	\tikzstyle{level 1}=[level distance=24pt]
    	\coordinate
    		child{ edge from parent [full edge]
    		} ;
        \end{tikzpicture} \quad &= \quad
    \begin{tikzpicture} [thick,scale=0.95]
    	\tikzstyle{level 1}=[level distance=24pt]
    	\coordinate
    		child{ edge from parent [dash edge]
    		} ;
    \end{tikzpicture} \quad = \quad
    \begin{tikzpicture} [thick,scale=0.95]
        \tikzstyle{level 1}=[level distance=24pt]
    	\coordinate
    		child{ edge from parent [dot edge]
    		} ;
    \end{tikzpicture}  \quad \triangleq 1 \quad , \quad
    \begin{tikzpicture} [thick,scale=0.5]
    	\tikzstyle{level 1}=[level distance=24pt]
    	\tikzstyle{level 2}=[level distance=24pt, sibling distance=24pt]
    	\coordinate
    		child{ edge from parent [dash edge]
    			child{ edge from parent [dash edge] }
    			child{ edge from parent [dash edge] }
    		} ;
        \end{tikzpicture} \triangleq x \quad , \quad 
    \begin{tikzpicture} [thick,scale=0.5]
    	\tikzstyle{level 1}=[level distance=24pt]
    	\tikzstyle{level 2}=[level distance=24pt, sibling distance=24pt]
    	\coordinate
    		child{ edge from parent [dot edge]
    			child{ edge from parent [dot edge] }
    			child{ edge from parent [dot edge] }
    		} ;
    \end{tikzpicture} \triangleq \xi \quad , \quad 
    \begin{tikzpicture} [thick,scale=0.5]
    	\tikzstyle{level 1}=[level distance=24pt]
    	\tikzstyle{level 2}=[level distance=24pt, sibling distance=24pt]
    	\coordinate
    		child{ edge from parent [full edge]
    			child{ edge from parent [full edge] }
    			child{ edge from parent [dash edge] }
    		} ;
    \end{tikzpicture} \triangleq z \quad , \quad 
    \begin{tikzpicture} [thick,scale=0.5]
    	\tikzstyle{level 1}=[level distance=24pt]
    	\tikzstyle{level 2}=[level distance=24pt, sibling distance=24pt]
    	\coordinate
    		child{ edge from parent [full edge]
    			child{ edge from parent [dot edge] }
    			child{ edge from parent [full edge] }
    		} ;
    \end{tikzpicture} \triangleq \zeta \quad ,
    \nonumber\\
    \begin{tikzpicture} [thick,scale=0.95]	
    	\tikzstyle{level 1}=[level distance=24pt]
    	\coordinate
        child{ edge from parent [dash edge] circle(3pt) } ;	
        \end{tikzpicture} &\triangleq {\gC}(x)\quad , \quad
    \begin{tikzpicture} [thick,scale=0.95]	
    	\tikzstyle{level 1}=[level distance=24pt]
    	\coordinate
    		child{ edge from parent [dot edge] circle(3pt) } ;	
    \end{tikzpicture} \triangleq {\gC}(\xi)\quad , \quad
    \begin{tikzpicture} [thick,scale=0.95]
    	\tikzstyle{level 1}=[level distance=24pt]
    	\coordinate
    		child{ edge from parent [dressed edge] } ;
    \end{tikzpicture} \triangleq {\gK}(x,z) \quad , \quad
    \begin{tikzpicture} [thick,scale=0.95]
    	\tikzstyle{level 1}=[level distance=24pt]
    	\coordinate
    		child{ edge from parent [redressed edge] } ;
        \end{tikzpicture} \triangleq {\gJ}(\xi,x,z,\zeta) ~.
\end{align}
In advance we can tell that anything happening to the right of the path can be neglected if we replace the edges in the path with external rooted paths. 
This lets us skip ahead a number of steps and write
\begin{align}
    \begin{tikzpicture} [thick,scale=0.5]	
    	\tikzstyle{level 1}=[level distance=24pt]
    	\coordinate
    			child{ edge from parent [redressed edge] } ;	
    \end{tikzpicture}
    ~~&=~~
    \begin{tikzpicture} [thick,scale=0.5]
    	\tikzstyle{level 1}=[level distance=24pt]
    	\coordinate
    		child{ edge from parent [dressed edge]
    		} ;
    \end{tikzpicture}
    ~~+~~
    \begin{tikzpicture} [thick,scale=0.5]
    	\tikzstyle{level 1}=[level distance=24pt]
    	\tikzstyle{level 2}=[level distance=24pt, sibling distance=24pt]
    	\coordinate
    		child{ edge from parent [dressed edge]
    			child{ edge from parent [dot edge] }
    			child{ edge from parent [dressed edge] }
    		} ;
    \end{tikzpicture}
    ~~+~~
    \begin{tikzpicture} [thick,scale=0.5]
    	\tikzstyle{level 1}=[level distance=24pt]
    	\tikzstyle{level 2}=[level distance=24pt, sibling distance=24pt]
    	\tikzstyle{level 3}=[level distance=24pt, sibling distance=24pt]
    	\coordinate
    		child{ edge from parent [dressed edge]
    			child{ edge from parent [dot edge]
    				child{ edge from parent [dot edge] }
    				child{ edge from parent [dot edge] }			
    			}
    			child{ edge from parent [dressed edge] }
    		} ;
    \end{tikzpicture}
    ~~+~~
    \begin{tikzpicture} [thick,scale=0.5]
    	\tikzstyle{level 1}=[level distance=24pt]
    	\tikzstyle{level 2}=[level distance=24pt, sibling distance=24pt]
    	\tikzstyle{level 3}=[level distance=24pt, sibling distance=24pt]
    	\coordinate
    		child{ edge from parent [dressed edge]
    			child{ edge from parent [dot edge] }
    			child{ edge from parent [dressed edge]
    				child{ edge from parent [dot edge] }
    				child{ edge from parent [dressed edge] }			
    			}
    		} ;
    \end{tikzpicture}
    ~~+~~
    \begin{tikzpicture} [thick,scale=0.5]
    	\tikzstyle{level 1}=[level distance=24pt]
    	\tikzstyle{level 2}=[level distance=24pt, sibling distance=24pt]
    	\tikzstyle{level 3}=[level distance=24pt, sibling distance=20pt]
    	\coordinate
    		child{ edge from parent [dressed edge]
    			child{ edge from parent [dot edge]
    				child{ edge from parent [dot edge] }
    				child{ edge from parent [dot edge] }			
    			}
    			child{ edge from parent [dressed edge]
    				child{ edge from parent [dot edge] }
    				child{ edge from parent [dressed edge] }			
    			}
    		} ;
    \end{tikzpicture}
    ~~+\dots ~.
\end{align}
With our experience so far we can sum out the $\xi$ vertices, denoting Catalan sub-trees, to get
\begin{align}
    \begin{tikzpicture} [thick,scale=0.5]	
    	\tikzstyle{level 1}=[level distance=24pt]
    	\coordinate
    			child{ edge from parent [redressed edge] } ;	
    \end{tikzpicture}
    ~~&=~~
    \begin{tikzpicture} [thick,scale=0.5]
    	\tikzstyle{level 1}=[level distance=24pt]
    	\coordinate
    		child{ edge from parent [dressed edge]
    		} ;
    \end{tikzpicture}
    ~~+~~
    \begin{tikzpicture} [thick,scale=0.5]
    	\tikzstyle{level 1}=[level distance=24pt]
    	\tikzstyle{level 2}=[level distance=24pt, sibling distance=24pt]
    	\coordinate
    		child{ edge from parent [dressed edge]
    			child{ edge from parent [dot edge]  circle(3pt)  }
    			child{ edge from parent [dressed edge] }
    		} ;
    \end{tikzpicture}
    ~~+~~
    \begin{tikzpicture} [thick,scale=0.5]
    	\tikzstyle{level 1}=[level distance=24pt]
    	\tikzstyle{level 2}=[level distance=24pt, sibling distance=24pt]
    	\tikzstyle{level 3}=[level distance=24pt, sibling distance=24pt]
    	\coordinate
    		child{ edge from parent [dressed edge]
    			child{ edge from parent [dot edge]   circle(3pt)  }
    			child{ edge from parent [dressed edge]
    				child{ edge from parent [dot edge]  circle(3pt)  }
    				child{ edge from parent [dressed edge] }			
    			}
    		} ;
    \end{tikzpicture}
    ~~+\dots ~,
\end{align}
so that we arrive at the recursion relation
\begin{align}
    \begin{tikzpicture} [thick,scale=0.5]	
    	\tikzstyle{level 1}=[level distance=24pt]
    	\coordinate
    			child{ edge from parent [redressed edge] } ;	
    \end{tikzpicture}
    ~~&=~~
    \begin{tikzpicture} [thick,scale=0.5]
    	\tikzstyle{level 1}=[level distance=24pt]
    	\coordinate
    		child{ edge from parent [dressed edge]
    		} ;
    \end{tikzpicture}
    ~~+~~
    \begin{tikzpicture} [thick,scale=0.5]
    	\tikzstyle{level 1}=[level distance=24pt]
    	\tikzstyle{level 2}=[level distance=24pt, sibling distance=24pt]
    	\coordinate
    		child{ edge from parent [dressed edge]
    			child{ edge from parent [dot edge]  circle(3pt)  }
    			child{ edge from parent [redressed edge] }
    		} ;
    \end{tikzpicture} ~.
\end{align}
We again translate back into algebra and find
\begin{align}
    {\gJ}(\xi,x,z,\zeta) &= {\gK}(x,z) + \zeta {\gC}(\xi) {\gK}(x,z) {\gJ}(\xi,x,z,\zeta) \\
    &= \frac{{\gK}(x,z)}{1-\zeta {\gC}(\xi) {\gK}(x,z) } 
    = \frac{1}{1-z{\gC}(x)-\zeta {\gC}(\xi)} ~.
    \label{eq-genRootPath}
\end{align}
For a path to have penetration $p$ we need to have $i$ vertices to the left of the path ($\xi$) and $l$ right turning vertices ($\zeta$) such that $i+l=p$. 
The way we enforce this is to multiply both $\xi$ and $\zeta$ by some dummy variable $y$ within the generating function and then look at the coefficient of $y^p$ in the resulting expansion. 
Let us therefore define
\begin{equation}
    \vJ^{(p)}(\xi,x,z,\zeta) = [y^p] \{ {\gJ}(y\xi,x,z,y\zeta)  \} 
                             = [y^p]\left\{\frac{1}{1-z{\gC}(x)-y\zeta {\gC}(y\xi)}\right\} ~.
    \label{eq-vJ}
\end{equation}
From this we can write down the sum of the depths of the $p^\text{th}$ leaf (starting from zero) as
\begin{equation}
    \vL^{(p)}_n = \sum_{i+j+k+l=n} (k+l)  [\xi^i x^j z^k \zeta^l] [y^p] \{ {\gJ}(y\xi,x,z,y\zeta)  \} ~.
\end{equation}
Exactly as before we can manipulate this expression to bring it into a simpler form using equations \eqref{eq-genfuneuler} and \eqref{eq-sumout}
\begin{equation}
    \vL^{(p)}_n = [x^n y^p]\left\{ \frac{1}{\left[ x{\gC}(x)+xy {\gC}(xy) -1\right] } + \frac{1}{\left[ x{\gC}(x)+xy {\gC}(xy) -1\right]^2 } \right\} ~.
    \label{eq-depthGF}
\end{equation}
This is the generating function found by Kirchenhoffer.\cite{Kir83}
The explicit form is   
\begin{equation}
    \vL^{(p)}_n = \frac{2(p+1)(2p+1)(n-p+1)(2n-2p+1)}{(n+1)(n+2)} \vC_{p} \vC_{n-p} - \vC_{n} ~,
    \label{eq-Lpn}
\end{equation}
(cp.\ appendix \ref{sec:DepthDeriv}).
The average depth of a leaf with penetration $p$
%, $\aL^{(p)}_n = \frac{\vL^1{(p)}_{n}}{\vC_{n}}$ 
is then
\begin{equation}
    \aL^{(p)}_n 
    = \frac{\vL^{(p)}_{n}}{\vC_{n}}
    = \frac{2(p+1)(2p+1)(n-p+1)(2n-2p+1)}{(n+1)(n+2)} \frac{\vC_{p} \vC_{n-p}}{\vC_{n}} - 1 ~.
    \label{eq-av_depth}
\end{equation}
We note that a derivation of \eqref{eq-av_depth} without the diagrammatic approach is also possible, albeit less adaptable \cite{GolFRB15}.

%%%%%%%%%%%%%%%%%%%%%%%%%%%%%%%%%%%%%%%%%%%%%%%%%%%%%%%%%%%%%%%%%%%
\section{Average leaf-to-leaf path length in an ordered Catalan tree}
%%%%%%%%%%%%%%%%%%%%%%%%%%%%%%%%%%%%%%%%%%%%%%%%%%%%%%%%%%%%%%%%%%%%
Our primary task is to calculate the average length of the path separating two leaves on the tree.
For convenience in the following derivation we will define the \emph{separation} $s$ as the number of leaves in between the two leaves under consideration.
We note that this is one less than the usual definition of distance on a lattice in condensed matter physics (i.e. $s=r-1$, where $r = |x_{2} - x_{1}|$ if $x$ is the position on the lattice). 
By this definition adjacent leaves have separation $s=0$.
In order to enumerate leaf-to-leaf paths with a given separation it is necessary to distinguish between the left and right arcs of the path following the initial bifurcation. 
Furthermore we need to distinguish between those vertices sitting between the two arcs of the path and those vertices sitting to the left of the left arc or the right of the right arc. 
For our purposes, those vertices to the left of the left arc need not be distinguished from those to the right of the right arc. 
The vertices within an arc need to be separated into those which turn left and those which turn right so our diagrammatic dictionary must contain the following vertices
\begin{subequations}
    \begin{align}
    \begin{tikzpicture} [thick,scale=0.95]
    	\tikzstyle{level 1}=[level distance=24pt]
    	\coordinate
    		child{ edge from parent [full edge]
    		} ;
        \end{tikzpicture} \quad &= \quad
    \begin{tikzpicture} [thick,scale=0.95]
    	\tikzstyle{level 1}=[level distance=24pt]
    	\coordinate
    		child{ edge from parent [double edge]
    		} ;
    \end{tikzpicture} \quad = \quad
    \begin{tikzpicture} [thick,scale=0.95]
    	\tikzstyle{level 1}=[level distance=24pt]
    	\coordinate
    		child{ edge from parent [dash edge]
    		} ;
    \end{tikzpicture} \quad = \quad
    \begin{tikzpicture} [thick,scale=0.95]
    	\tikzstyle{level 1}=[level distance=24pt]
    	\coordinate
    		child{ edge from parent [dot edge]
    		} ;
    \end{tikzpicture}  \quad \triangleq 1 \: , \qquad 
    \begin{tikzpicture} [thick,scale=0.5]
    	\tikzstyle{level 1}=[level distance=24pt]
    	\tikzstyle{level 2}=[level distance=24pt, sibling distance=24pt]
    	\coordinate
    		child{ edge from parent [dash edge]
    			child{ edge from parent [dash edge] }
    			child{ edge from parent [dash edge] }
    		} ;
    \end{tikzpicture} \triangleq x \: , \qquad
    \begin{tikzpicture} [thick,scale=0.5]
    	\tikzstyle{level 1}=[level distance=24pt]
    	\tikzstyle{level 2}=[level distance=24pt, sibling distance=24pt]
    	\coordinate
    		child{ edge from parent [dot edge]
    			child{ edge from parent [dot edge] }
    			child{ edge from parent [dot edge] }
    		} ;
    \end{tikzpicture} \triangleq \xi \: , \qquad
    \nonumber\\
    &\quad
    \begin{tikzpicture} [thick,scale=0.5]
    	\tikzstyle{level 1}=[level distance=24pt]
    	\tikzstyle{level 2}=[level distance=24pt, sibling distance=24pt]
    	\coordinate
    		child{ edge from parent [full edge]
    			child{ edge from parent [full edge] }
    			child{ edge from parent [dash edge] }
    		} ;
    \end{tikzpicture} \triangleq z \: , \qquad 
    \begin{tikzpicture} [thick,scale=0.5]
    	\tikzstyle{level 1}=[level distance=24pt]
    	\tikzstyle{level 2}=[level distance=24pt, sibling distance=24pt]
    	\coordinate
    		child{ edge from parent [full edge]
    			child{ edge from parent [dot edge] }
    			child{ edge from parent [full edge] }
    		} ;
    \end{tikzpicture} \triangleq \zeta \: , \qquad 
    \begin{tikzpicture} [thick,scale=0.5]
    	\tikzstyle{level 1}=[level distance=24pt]
    	\tikzstyle{level 2}=[level distance=24pt, sibling distance=24pt]
    	\coordinate
    		child{ edge from parent [double edge]
    			child{ edge from parent [dash edge] }
    			child{ edge from parent [double edge] }
    		} ;
    \end{tikzpicture} \triangleq \bar{z} \: , \qquad
    \begin{tikzpicture} [thick,scale=0.5]
    	\tikzstyle{level 1}=[level distance=24pt]
    	\tikzstyle{level 2}=[level distance=24pt, sibling distance=24pt]
    	\coordinate
    		child{ edge from parent [double edge]
    			child{ edge from parent [double edge] }
    			child{ edge from parent [dot edge] }
    		} ;
    \end{tikzpicture} \triangleq \bar{\zeta} ~.
\end{align}
As yet we have made no reference to those vertices sitting above/outside the initial bifurcation. To account for these we need to extend our diagrammatic dictionary to incorporate the following
\begin{equation}
    \begin{tikzpicture} [thick,scale=0.95]
    	\tikzstyle{level 1}=[level distance=24pt]
    	\coordinate
    		child{ edge from parent [snake edge]
    		} ;
    \end{tikzpicture}  \quad \triangleq 1 \quad , \qquad
    \begin{tikzpicture} [thick,scale=0.5]
    	\tikzstyle{level 1}=[level distance=24pt]
    	\tikzstyle{level 2}=[level distance=24pt, sibling distance=24pt]
    	\coordinate
    		child{ edge from parent [snake edge]
    			child{ edge from parent [snake edge] }
    			child{ edge from parent [snake edge] }
    		} ;
    \end{tikzpicture} \triangleq \lambda \quad , \qquad
    \begin{tikzpicture} [thick,scale=0.5]
    	\tikzstyle{level 1}=[level distance=24pt]
    	\tikzstyle{level 2}=[level distance=24pt, sibling distance=24pt]
    	\coordinate
    		child{ edge from parent [snake edge]
    			child{ edge from parent [full edge] }
    			child{ edge from parent [double edge] }
    		} ;
    \end{tikzpicture} \triangleq \mu ~.
\end{equation}
\end{subequations}
The generating function for leaf-to-leaf paths needs to enumerate those trees containing
(i) $i$ vertices to the left of the left arc or the right of the right arc,
(ii) $j$ vertices to the right of the left arc and the left of the right arc,
(iii) $k$ left-turning vertices within the left arc,
(iv) $l$ right-turning vertices within the left arc,
(v) $\bar{k}$ right-turning vertices within the right arc,
(vi) $\bar{l}$ left-turning vertices within the right arc,
(vii) $m$ vertices above the initial bifurcation,
(viii) $1$ initial bifurcation.
The series so generated will be denoted
\begin{equation}
    \vM_{ijkl\bar{k}\bar{l}m}= [\xi^i x^j z^k \zeta^l \bar{z}^{\bar{k}} \bar{\zeta}^{\bar{l}} \lambda^m \mu] \left\{ \mu {\gM}(\xi,x,z,\zeta,\bar{z},\bar{\zeta},\lambda) \right\} ~.
\end{equation}
We will represent the generating function with the diagram
\begin{equation}
    \begin{tikzpicture}[scale=0.5,thick]	
    	\tikzstyle{level 1}=[level distance=24pt]
    	\coordinate
    			child{ edge from parent [snake edge] [fill=black] circle (8pt)
    				child { edge from parent [full edge] }
    				child { edge from parent [double edge] }			
    			} ;	
    \end{tikzpicture} \quad \triangleq \quad \mu {\gM}(\xi,x,z,\zeta,\bar{z},\bar{\zeta},\lambda) ~.
\end{equation}
We begin by writing out the diagrammatic series
\begin{align}
    \begin{tikzpicture}[scale=0.4,thick]	
    	\tikzstyle{level 1}=[level distance=24pt]
    	\coordinate
    			child{ edge from parent [snake edge] [fill=black] circle (8pt)
    				child { edge from parent [full edge] }
    				child { edge from parent [double edge] }			
    			} ;	
    \end{tikzpicture}
    &~=~
    \begin{tikzpicture}[scale=0.4,thick]	
    	\tikzstyle{level 1}=[level distance=24pt]
    	\coordinate
    			child{ edge from parent [snake edge]
    				child { edge from parent [full edge] }
    				child { edge from parent [double edge] }			
    			} ;	
    \end{tikzpicture}
    ~+~
    \begin{tikzpicture}[scale=0.4,thick]	
    	\tikzstyle{level 1}=[level distance=24pt]
    	\coordinate
    			child{ edge from parent [snake edge]
    				child { edge from parent [snake edge]
    					child { edge from parent [full edge] }
    					child { edge from parent [double edge] }				
    				}
    				child { edge from parent [snake edge] }
    			} ;	
    \end{tikzpicture}
    ~+~
    \begin{tikzpicture}[scale=0.4,thick]	
    	\tikzstyle{level 1}=[level distance=24pt]
    	\coordinate
    			child{ edge from parent [snake edge]
    				child { edge from parent [full edge]
    					child { edge from parent [full edge] }
    					child { edge from parent [dash edge] }				
    				}
    				child { edge from parent [double edge] }			
    			} ;	
    \end{tikzpicture}
    ~+~
    \begin{tikzpicture}[scale=0.4,thick]	
    	\tikzstyle{level 1}=[level distance=24pt]
    	\coordinate
    			child{ edge from parent [snake edge]
    				child { edge from parent [full edge]
    					child { edge from parent [dot edge] }
    					child { edge from parent [full edge] }				
    				}
    				child { edge from parent [double edge] }			
    			} ;	
    \end{tikzpicture}
    ~+~
    \begin{tikzpicture}[scale=0.4,thick]	
    	\tikzstyle{level 1}=[level distance=24pt]
    	\coordinate
    			child{ edge from parent [snake edge]
    				child { edge from parent [snake edge] }
    				child { edge from parent [snake edge]
    					child { edge from parent [full edge] }
    					child { edge from parent [double edge] }				
    				}
    			} ;	
    \end{tikzpicture}
%    \nonumber 
%    %%%new line%%%%%%%%
%    \\
%    &
    ~+~
    \begin{tikzpicture}[scale=0.4,thick]	
    	\tikzstyle{level 1}=[level distance=24pt]
    	\coordinate
    			child{ edge from parent [snake edge]
    				child { edge from parent [full edge] }
    				child { edge from parent [double edge]
    					child { edge from parent [dash edge] }
    					child { edge from parent [double edge] }				
    				}
    			} ;	
    \end{tikzpicture}
    ~+~
    \begin{tikzpicture}[scale=0.4,thick]	
    	\tikzstyle{level 1}=[level distance=24pt]
    	\coordinate
    			child{ edge from parent [snake edge]
    				child { edge from parent [full edge] }
    				child { edge from parent [double edge]
    					child { edge from parent [double edge] }
    					child { edge from parent [dot edge] }				
    				}
    			} ;	
    \end{tikzpicture}
    ~+~
    \begin{tikzpicture}[scale=0.4,thick]	
    	\tikzstyle{level 1}=[level distance=24pt]
    	\tikzstyle{level 2}=[level distance=24pt]
    	\tikzstyle{level 3}=[level distance=24pt, sibling distance=24pt]
    	\coordinate
    			child{ edge from parent [snake edge]
    				child { edge from parent [snake edge]
    					child { edge from parent [full edge] }
    					child { edge from parent [double edge] }
    				}
    				child { edge from parent [snake edge]
    					child { edge from parent [snake edge] }
    					child { edge from parent [snake edge] }				
    				}
    			} ;	
    \end{tikzpicture}
    \nonumber 
    %%%new line%%%%%%%%
    \\
    &\quad
    ~+~
    \begin{tikzpicture}[scale=0.4,thick]	
    	\tikzstyle{level 1}=[level distance=24pt]
    	\tikzstyle{level 2}=[level distance=24pt]
    	\tikzstyle{level 3}=[level distance=24pt, sibling distance=24pt]
    	\coordinate
    			child{ edge from parent [snake edge]
    				child { edge from parent [snake edge]
    					child { edge from parent [snake edge] }
    					child { edge from parent [snake edge] }
    				}
    				child { edge from parent [snake edge]
    					child { edge from parent [full edge] }
    					child { edge from parent [double edge] }				
    				}
    			} ;	
    \end{tikzpicture}
    ~+~
    \begin{tikzpicture}[scale=0.4,thick]	
    	\tikzstyle{level 1}=[level distance=24pt]
    	\tikzstyle{level 2}=[level distance=24pt]
    	\tikzstyle{level 3}=[level distance=24pt, sibling distance=24pt]
    	\coordinate
    			child{ edge from parent [snake edge]
    				child { edge from parent [full edge]
    					child { edge from parent [dot edge] }
    					child { edge from parent [full edge] }
    				}
    				child { edge from parent [double edge]
    					child { edge from parent [double edge] }
    					child { edge from parent [dot edge] }				
    				}
    			} ;	
    \end{tikzpicture}
    ~+~
    \begin{tikzpicture}[scale=0.4,thick]	
    	\tikzstyle{level 1}=[level distance=24pt]
    	\tikzstyle{level 2}=[level distance=24pt]
    	\tikzstyle{level 3}=[level distance=24pt, sibling distance=24pt]
    	\coordinate
    			child{ edge from parent [snake edge]
    				child { edge from parent [full edge]
    					child { edge from parent [dot edge] }
    					child { edge from parent [full edge] }
    				}
    				child { edge from parent [double edge]
    					child { edge from parent [dash edge] }
    					child { edge from parent [double edge] }				
    				}
    			} ;	
    \end{tikzpicture}
    ~+~
    \begin{tikzpicture}[scale=0.4,thick]	
    	\tikzstyle{level 1}=[level distance=24pt]
    	\tikzstyle{level 2}=[level distance=24pt]
    	\tikzstyle{level 3}=[level distance=24pt, sibling distance=24pt]
    	\coordinate
    			child{ edge from parent [snake edge]
    				child { edge from parent [full edge]
    					child { edge from parent [full edge] }
    					child { edge from parent [dash edge] }
    				}
    				child { edge from parent [double edge]
    					child { edge from parent [double edge] }
    					child { edge from parent [dot edge] }				
    				}
    			} ;	
    \end{tikzpicture}
    ~+~
    \begin{tikzpicture}[scale=0.4,thick]	
    	\tikzstyle{level 1}=[level distance=24pt]
    	\tikzstyle{level 2}=[level distance=24pt]
    	\tikzstyle{level 3}=[level distance=24pt, sibling distance=24pt]
    	\coordinate
    			child{ edge from parent [snake edge]
    				child { edge from parent [full edge]
    					child { edge from parent [full edge] }
    					child { edge from parent [dash edge] }
    				}
    				child { edge from parent [double edge]
    					child { edge from parent [dash edge] }
    					child { edge from parent [double edge] }				
    				}
    			} ;	
    \end{tikzpicture}
    ~+~
    \begin{tikzpicture}[scale=0.3,thick]	
    	\tikzstyle{level 1}=[level distance=24pt]
    	\tikzstyle{level 2}=[level distance=24pt]
    	\tikzstyle{level 3}=[level distance=24pt]
    	\coordinate
    			child{ edge from parent [snake edge]
    				child { edge from parent [snake edge]
    					child { edge from parent [snake edge]
    						child { edge from parent [full edge] }
    						child { edge from parent [double edge] }
    					}
    					child { edge from parent [snake edge] }
    				}
    				child { edge from parent [snake edge] }
    			} ;	
    \end{tikzpicture}
    ~+~
    \begin{tikzpicture}[scale=0.3,thick]	
    	\tikzstyle{level 1}=[level distance=24pt]
    	\tikzstyle{level 2}=[level distance=24pt]
    	\tikzstyle{level 3}=[level distance=24pt]
    	\coordinate
    			child{ edge from parent [snake edge]
    				child { edge from parent [snake edge]
    					child { edge from parent [full edge]
    						child { edge from parent [full edge] }
    						child { edge from parent [dash edge] }
    					}
    					child { edge from parent [double edge] }
    				}
    				child { edge from parent [snake edge] }
    			} ;	
    \end{tikzpicture}
    ~+~
    \begin{tikzpicture}[scale=0.3,thick]	
    	\tikzstyle{level 1}=[level distance=24pt]
    	\tikzstyle{level 2}=[level distance=24pt]
    	\tikzstyle{level 3}=[level distance=24pt]
    	\coordinate
    			child{ edge from parent [snake edge]
    				child { edge from parent [snake edge]
    					child { edge from parent [full edge]
    						child { edge from parent [dot edge] }
    						child { edge from parent [full edge] }
    					}
    					child { edge from parent [double edge] }
    				}
    				child { edge from parent [snake edge] }
    			} ;	
    \end{tikzpicture}
    \nonumber 
    %%%new line%%%%%%%%
    \\
    &\quad
    ~+~
    \begin{tikzpicture}[scale=0.3,thick]	
    	\tikzstyle{level 1}=[level distance=24pt]
    	\tikzstyle{level 2}=[level distance=24pt]
    	\tikzstyle{level 3}=[level distance=24pt]
    	\coordinate
    			child{ edge from parent [snake edge]
    				child { edge from parent [full edge]
    					child { edge from parent [full edge]
    						child { edge from parent [full edge] }
    						child { edge from parent [dash edge] }
    					}
    					child { edge from parent [dash edge] }
    				}
    				child { edge from parent [double edge] }
    			} ;	
    \end{tikzpicture}
    ~+~
    \begin{tikzpicture}[scale=0.3,thick]	
    	\tikzstyle{level 1}=[level distance=24pt]
    	\tikzstyle{level 2}=[level distance=24pt]
    	\tikzstyle{level 3}=[level distance=24pt]
    	\coordinate
    			child{ edge from parent [snake edge]
    				child { edge from parent [full edge]
    					child { edge from parent [full edge]
    						child { edge from parent [dot edge] }
    						child { edge from parent [full edge] }
    					}
    					child { edge from parent [dash edge] }
    				}
    				child { edge from parent [double edge] }
    			} ;	
    \end{tikzpicture}
    ~+~
    \begin{tikzpicture}[scale=0.3,thick]	
    	\tikzstyle{level 1}=[level distance=24pt]
    	\tikzstyle{level 2}=[level distance=24pt]
    	\tikzstyle{level 3}=[level distance=24pt]
    	\coordinate
    			child{ edge from parent [snake edge]
    				child { edge from parent [full edge]
    					child { edge from parent [dot edge]
    						child { edge from parent [dot edge] }
    						child { edge from parent [dot edge] }
    					}
    					child { edge from parent [full edge] }
    				}
    				child { edge from parent [double edge] }
    			} ;	
    \end{tikzpicture}
    ~+~
    \begin{tikzpicture}[scale=0.3,thick]	
    	\tikzstyle{level 1}=[level distance=24pt]
    	\tikzstyle{level 2}=[level distance=24pt]
    	\tikzstyle{level 3}=[level distance=24pt]
    	\coordinate
    			child{ edge from parent [snake edge]
    				child { edge from parent [snake edge] }
    				child { edge from parent [snake edge]
    					child { edge from parent [snake edge] }
    					child { edge from parent [snake edge]
    						child { edge from parent [full edge] }
    						child { edge from parent [double edge] }
    					}
    				}
    			} ;	
    \end{tikzpicture}
    ~+~
    \begin{tikzpicture}[scale=0.3,thick]	
    	\tikzstyle{level 1}=[level distance=24pt]
    	\tikzstyle{level 2}=[level distance=24pt]
    	\tikzstyle{level 3}=[level distance=24pt]
    	\coordinate
    			child{ edge from parent [snake edge]
    				child { edge from parent [snake edge] }
    				child { edge from parent [snake edge]
    					child { edge from parent [full edge] }
    					child { edge from parent [double edge]
    						child { edge from parent [dash edge] }
    						child { edge from parent [double edge] }
    					}
    				}
    			} ;	
    \end{tikzpicture}
    \nonumber 
    ~+~
    \begin{tikzpicture}[scale=0.3,thick]	
    	\tikzstyle{level 1}=[level distance=24pt]
    	\tikzstyle{level 2}=[level distance=24pt]
    	\tikzstyle{level 3}=[level distance=24pt]
    	\coordinate
    			child{ edge from parent [snake edge]
    				child { edge from parent [snake edge] }
    				child { edge from parent [snake edge]
    					child { edge from parent [full edge] }
    					child { edge from parent [double edge]
    						child { edge from parent [double edge] }
    						child { edge from parent [dot edge] }
    					}
    				}
    			} ;	
    \end{tikzpicture}
    ~+~
    \begin{tikzpicture}[scale=0.3,thick]	
    	\tikzstyle{level 1}=[level distance=24pt]
    	\tikzstyle{level 2}=[level distance=24pt]
    	\tikzstyle{level 3}=[level distance=24pt]
    	\coordinate
    			child{ edge from parent [snake edge]
    				child { edge from parent [full edge] }
    				child { edge from parent [double edge]
    					child { edge from parent [dash edge] }
    					child { edge from parent [double edge]
    						child { edge from parent [dash edge] }
    						child { edge from parent [double edge] }
    					}
    				}
    			} ;	
    \end{tikzpicture}
    ~+~
    \begin{tikzpicture}[scale=0.3,thick]	
    	\tikzstyle{level 1}=[level distance=24pt]
    	\tikzstyle{level 2}=[level distance=24pt]
    	\tikzstyle{level 3}=[level distance=24pt]
    	\coordinate
    			child{ edge from parent [snake edge]
    				child { edge from parent [full edge] }
    				child { edge from parent [double edge]
    					child { edge from parent [dash edge] }
    					child { edge from parent [double edge]
    						child { edge from parent [double edge] }
    						child { edge from parent [dot edge] }
    					}
    				}
    			} ;	
    \end{tikzpicture}
    %%%new line%%%%%%%%
    \nonumber 
    \\
    &\quad
    ~+~
    \begin{tikzpicture}[scale=0.3,thick]	
    	\tikzstyle{level 1}=[level distance=24pt]
    	\tikzstyle{level 2}=[level distance=24pt]
    	\tikzstyle{level 3}=[level distance=24pt]
    	\coordinate
    			child{ edge from parent [snake edge]
    				child { edge from parent [full edge] }
    				child { edge from parent [double edge]
    					child { edge from parent [double edge] }
    					child { edge from parent [dot edge]
    						child { edge from parent [dot edge] }
    						child { edge from parent [dot edge] }
    					}
    				}
    			} ;	
    \end{tikzpicture}
    ~+~~
    \begin{tikzpicture}[scale=0.3,thick]	
    	\tikzstyle{level 1}=[level distance=24pt]
    	\tikzstyle{level 2}=[level distance=24pt]
    	\tikzstyle{level 3}=[level distance=24pt]
    	\coordinate
    			child{ edge from parent [snake edge]
    				child { edge from parent [snake edge]
    					child { edge from parent [snake edge] }
    					child { edge from parent [snake edge]
    						child { edge from parent [full edge] }
    						child { edge from parent [double edge] }
    					}
    				}
    				child { edge from parent [snake edge] }
    			} ;	
    \end{tikzpicture}
    ~~+~~
    \begin{tikzpicture}[scale=0.3,thick]	
    	\tikzstyle{level 1}=[level distance=24pt]
    	\tikzstyle{level 2}=[level distance=24pt]
    	\tikzstyle{level 3}=[level distance=24pt]
    	\coordinate
    			child{ edge from parent [snake edge]
    				child { edge from parent [snake edge]
    					child { edge from parent [full edge] }
    					child { edge from parent [double edge]
    						child { edge from parent [dash edge] }
    						child { edge from parent [double edge] }
    					}
    				}
    				child { edge from parent [snake edge] }
    			} ;	
    \end{tikzpicture}
    ~~+~~
    \begin{tikzpicture}[scale=0.3,thick]	
    	\tikzstyle{level 1}=[level distance=24pt]
    	\tikzstyle{level 2}=[level distance=24pt]
    	\tikzstyle{level 3}=[level distance=24pt]
    	\coordinate
    			child{ edge from parent [snake edge]
    				child { edge from parent [snake edge]
    					child { edge from parent [full edge] }
    					child { edge from parent [double edge]
    						child { edge from parent [double edge] }
    						child { edge from parent [dot edge] }
    					}
    				}
    				child { edge from parent [snake edge] }
    			} ;	
    \end{tikzpicture}
    ~~+~~
    \begin{tikzpicture}[scale=0.3,thick]	
    	\tikzstyle{level 1}=[level distance=24pt]
    	\tikzstyle{level 2}=[level distance=24pt]
    	\tikzstyle{level 3}=[level distance=24pt]
    	\coordinate
    			child{ edge from parent [snake edge]
    				child { edge from parent [full edge]
    					child { edge from parent [full edge] }
    					child { edge from parent [dash edge]
    						child { edge from parent [dash edge] }
    						child { edge from parent [dash edge] }
    					}
    				}
    				child { edge from parent [double edge] }
    			} ;	
    \end{tikzpicture}
    ~~+~~
    \begin{tikzpicture}[scale=0.3,thick]	
    	\tikzstyle{level 1}=[level distance=24pt]
    	\tikzstyle{level 2}=[level distance=24pt]
    	\tikzstyle{level 3}=[level distance=24pt]
    	\coordinate
    			child{ edge from parent [snake edge]
    				child { edge from parent [full edge]
    					child { edge from parent [dot edge] }
    					child { edge from parent [full edge]
    						child { edge from parent [full edge] }
    						child { edge from parent [dash edge] }
    					}
    				}
    				child { edge from parent [double edge] }
    			} ;	
    \end{tikzpicture}
    ~+~~
    \begin{tikzpicture}[scale=0.3,thick]	
    	\tikzstyle{level 1}=[level distance=24pt]
    	\tikzstyle{level 2}=[level distance=24pt]
    	\tikzstyle{level 3}=[level distance=24pt]
    	\coordinate
    			child{ edge from parent [snake edge]
    				child { edge from parent [full edge]
    					child { edge from parent [dot edge] }
    					child { edge from parent [full edge]
    						child { edge from parent [dot edge] }
    						child { edge from parent [full edge] }
    					}
    				}
    				child { edge from parent [double edge] }
    			} ;	
    \end{tikzpicture}
    ~~+~~
    \begin{tikzpicture}[scale=0.3,thick]	
    	\tikzstyle{level 1}=[level distance=24pt]
    	\tikzstyle{level 2}=[level distance=24pt]
    	\tikzstyle{level 3}=[level distance=24pt]
    	\coordinate
    			child{ edge from parent [snake edge]
    				child { edge from parent [snake edge] }
    				child { edge from parent [snake edge]
    					child { edge from parent [snake edge]
    						child { edge from parent [full edge] }
    						child { edge from parent [double edge] }
    					}
    					child { edge from parent [snake edge] }
    				}
    			} ;	
    \end{tikzpicture}
    %%%new line%%%%%%%%
    \nonumber 
    \\
    &\quad
    ~+~~
    \begin{tikzpicture}[scale=0.3,thick]	
    	\tikzstyle{level 1}=[level distance=24pt]
    	\tikzstyle{level 2}=[level distance=24pt]
    	\tikzstyle{level 3}=[level distance=24pt]
    	\coordinate
    			child{ edge from parent [snake edge]
    				child { edge from parent [snake edge] }
    				child { edge from parent [snake edge]
    					child { edge from parent [full edge]
    						child { edge from parent [full edge] }
    						child { edge from parent [dash edge] }
    					}
    					child { edge from parent [double edge] }
    				}
    			} ;	
    \end{tikzpicture}
    ~~+~~
    \begin{tikzpicture}[scale=0.3,thick]	
    	\tikzstyle{level 1}=[level distance=24pt]
    	\tikzstyle{level 2}=[level distance=24pt]
    	\tikzstyle{level 3}=[level distance=24pt]
    	\coordinate
    			child{ edge from parent [snake edge]
    				child { edge from parent [snake edge] }
    				child { edge from parent [snake edge]
    					child { edge from parent [full edge]
    						child { edge from parent [dot edge] }
    						child { edge from parent [full edge] }
    					}
    					child { edge from parent [double edge] }
    				}
    			} ;	
    \end{tikzpicture}
    ~~+~~
    \begin{tikzpicture}[scale=0.3,thick]	
    	\tikzstyle{level 1}=[level distance=24pt]
    	\tikzstyle{level 2}=[level distance=24pt]
    	\tikzstyle{level 3}=[level distance=24pt]
    	\coordinate
    			child{ edge from parent [snake edge]
    				child { edge from parent [full edge] }
    				child { edge from parent [double edge]
    					child { edge from parent [dash edge]
    						child { edge from parent [dash edge] }
    						child { edge from parent [dash edge] }
    					}
    					child { edge from parent [double edge] }
    				}
    			} ;	
    \end{tikzpicture}
    ~~+~~
    \begin{tikzpicture}[scale=0.3,thick]	
    	\tikzstyle{level 1}=[level distance=24pt]
    	\tikzstyle{level 2}=[level distance=24pt]
    	\tikzstyle{level 3}=[level distance=24pt]
    	\coordinate
    			child{ edge from parent [snake edge]
    				child { edge from parent [full edge] }
    				child { edge from parent [double edge]
    					child { edge from parent [double edge]
    						child { edge from parent [dash edge] }
    						child { edge from parent [double edge] }
    					}
    					child { edge from parent [dot edge] }
    				}
    			} ;	
    \end{tikzpicture}
    ~~+~~
    \begin{tikzpicture}[scale=0.3,thick]	
    	\tikzstyle{level 1}=[level distance=24pt]
    	\tikzstyle{level 2}=[level distance=24pt]
    	\tikzstyle{level 3}=[level distance=24pt]
    	\coordinate
    			child{ edge from parent [snake edge]
    				child { edge from parent [full edge] }
    				child { edge from parent [double edge]
    					child { edge from parent [double edge]
    						child { edge from parent [double edge] }
    						child { edge from parent [dot edge] }
    					}
    					child { edge from parent [dot edge] }
    				}
    			} ;	
    \end{tikzpicture}
    ~~+\dots ~.
\end{align}
We notice that certain diagrams within this series translate to the same term, for example
\begin{equation}
    \begin{tikzpicture}[scale=0.4,thick]	
    	\tikzstyle{level 1}=[level distance=24pt]
    	\coordinate
    			child{ edge from parent [snake edge]
    				child { edge from parent [snake edge]
    					child { edge from parent [full edge] }
    					child { edge from parent [double edge] }				
    				}
    				child { edge from parent [snake edge] }
    			} ;	
    \end{tikzpicture}
    ~~~ = ~~~
    \begin{tikzpicture}[scale=0.4,thick]	
    	\tikzstyle{level 1}=[level distance=24pt]
    	\coordinate
    			child{ edge from parent [snake edge]
    				child { edge from parent [snake edge] }
    				child { edge from parent [snake edge]
    					child { edge from parent [full edge] }
    					child { edge from parent [double edge] }				
    				}
    			} ;	
    \end{tikzpicture} ~,
\end{equation}
so that we ought to write
\begin{equation}
    \begin{tikzpicture}[scale=0.4,thick]	
    	\tikzstyle{level 1}=[level distance=24pt]
    	\coordinate
    			child{ edge from parent [snake edge]
    				child { edge from parent [snake edge]
    					child { edge from parent [full edge] }
    					child { edge from parent [double edge] }				
    				}
    				child { edge from parent [snake edge] }
    			} ;	
    \end{tikzpicture}
    ~~ + ~~
    \begin{tikzpicture}[scale=0.4,thick]	
    	\tikzstyle{level 1}=[level distance=24pt]
    	\coordinate
    			child{ edge from parent [snake edge]
    				child { edge from parent [snake edge] }
    				child { edge from parent [snake edge]
    					child { edge from parent [full edge] }
    					child { edge from parent [double edge] }				
    				}
    			} ;	
    \end{tikzpicture}
    ~~ = ~~ 2~
    \begin{tikzpicture}[scale=0.4,thick]	
    	\tikzstyle{level 1}=[level distance=24pt]
    	\coordinate
    			child{ edge from parent [snake edge]
    				child { edge from parent [snake edge] }
    				child { edge from parent [snake edge] }
    			} ;	
    \end{tikzpicture}
    ~~ \times ~~
    \begin{tikzpicture}[scale=0.4,thick]	
    	\tikzstyle{level 1}=[level distance=24pt]
    	\coordinate
    			child{ edge from parent [snake edge]
    				child { edge from parent [full edge] }
    				child { edge from parent [double edge] }
    			} ;	
    \end{tikzpicture} ~.
\end{equation}
Taking this into account, we factorise the generating function such that
\begin{align}
    \begin{tikzpicture}[scale=0.3,thick]	
    	\tikzstyle{level 1}=[level distance=24pt]
    	\coordinate
    			child{ edge from parent [snake edge] [fill=black] circle (8pt)
    				child { edge from parent [full edge] }
    				child { edge from parent [double edge] }			
    			} ;	
    \end{tikzpicture}
    ~&=
    \left(~
    \begin{tikzpicture}[scale=0.3,thick]	
    	\tikzstyle{level 1}=[level distance=24pt]
    	\coordinate
    			child{ edge from parent [snake edge] } ;	
    \end{tikzpicture}
    ~ + ~ 2 ~
    \begin{tikzpicture}[scale=0.3,thick]	
    	\tikzstyle{level 1}=[level distance=24pt]
    	\coordinate
    			child{ edge from parent [snake edge]
    				child { edge from parent [snake edge] }
    				child { edge from parent [snake edge] }			
    			} ;	
    \end{tikzpicture}
    ~ + ~ 3 ~
    \begin{tikzpicture}[scale=0.3,thick]	
    	\tikzstyle{level 1}=[level distance=24pt]
    	\coordinate
    			child{ edge from parent [snake edge]
    				child { edge from parent [snake edge]
    					child { edge from parent [snake edge] }
    					child { edge from parent [snake edge] }	
    				}
    				child { edge from parent [snake edge] }			
    			} ;	
    \end{tikzpicture}
    ~ + ~ 3 ~
    \begin{tikzpicture}[scale=0.3,thick]	
    	\tikzstyle{level 1}=[level distance=24pt]
    	\coordinate
    			child{ edge from parent [snake edge]
    				child { edge from parent [snake edge] }
    				child { edge from parent [snake edge]
    					child { edge from parent [snake edge] }
    					child { edge from parent [snake edge] }	
    				}			
    			} ;	
    \end{tikzpicture}
    ~ + ~ 4 ~
    \begin{tikzpicture}[scale=0.3,thick]	
    	\tikzstyle{level 1}=[level distance=24pt, sibling distance=24pt]
    	\tikzstyle{level 2}=[level distance=24pt, sibling distance=24pt]
    	\tikzstyle{level 3}=[level distance=24pt, sibling distance=20pt]
    	\coordinate
    			child{ edge from parent [snake edge]
    				child { edge from parent [snake edge]
    					child { edge from parent [snake edge] }
    					child { edge from parent [snake edge] }
    				}
    				child { edge from parent [snake edge]
    					child { edge from parent [snake edge] }
    					child { edge from parent [snake edge] }	
    				}			
    			} ;	
    \end{tikzpicture}
    ~ + ~ \dots \right) 
    %%%new line%%%%%%%%
    \nonumber \\
    & ~~~
    \times \left(~
    \begin{tikzpicture}[scale=0.3,thick]	
    	\tikzstyle{level 1}=[level distance=24pt]
    	\coordinate
    			child{ edge from parent [snake edge]
    				child { edge from parent [full edge] }
    				child { edge from parent [double edge] }			
    			} ;	
    \end{tikzpicture}
    ~ + ~
    \begin{tikzpicture}[scale=0.3,thick]	
    	\tikzstyle{level 1}=[level distance=24pt]
    	\coordinate
    			child{ edge from parent [snake edge]
    				child { edge from parent [full edge]
    					child { edge from parent [full edge]  }
    					child { edge from parent [dash edge]  }
    				}
    				child { edge from parent [double edge] }			
    			} ;	
    \end{tikzpicture}
    ~ + ~
    \begin{tikzpicture}[scale=0.3,thick]	
    	\tikzstyle{level 1}=[level distance=24pt]
    	\coordinate
    			child{ edge from parent [snake edge]
    				child { edge from parent [full edge]
    					child { edge from parent [dot edge]  }
    					child { edge from parent [full edge]  }
    				}
    				child { edge from parent [double edge] }			
    			} ;	
    \end{tikzpicture}
    ~ + ~
    \begin{tikzpicture}[scale=0.3,thick]	
    	\tikzstyle{level 1}=[level distance=24pt]
    	\coordinate
    			child{ edge from parent [snake edge]
    				child { edge from parent [full edge] }
    				child { edge from parent [double edge]
    					child { edge from parent [dash edge]  }
    					child { edge from parent [full edge]  }
    				}			
    			} ;	
    \end{tikzpicture}
    ~ + ~
    \begin{tikzpicture}[scale=0.3,thick]	
    	\tikzstyle{level 1}=[level distance=24pt]
    	\coordinate
    			child{ edge from parent [snake edge]
    				child { edge from parent [full edge] }
    				child { edge from parent [double edge]
    					child { edge from parent [full edge]  }
    					child { edge from parent [dot edge]  }
    				}			
    			} ;	
    \end{tikzpicture}
    ~ + ~ \dots
    ~ \right) ~.
\end{align}
The second series in parentheses can itself be factorised to give
\begin{align}
    \begin{tikzpicture}[scale=0.3,thick]	
    	\tikzstyle{level 1}=[level distance=24pt]
    	\coordinate
    			child{ edge from parent [snake edge] [fill=black] circle (8pt)
    				child { edge from parent [full edge] }
    				child { edge from parent [double edge] }			
    			} ;	
    \end{tikzpicture}
    ~&=
    \left( 
    \begin{tikzpicture}[scale=0.3,thick]	
    	\tikzstyle{level 1}=[level distance=24pt]
    	\coordinate
    			child{ edge from parent [snake edge] } ;	
    \end{tikzpicture}
    ~ + 2 ~
    \begin{tikzpicture}[scale=0.3,thick]	
    	\tikzstyle{level 1}=[level distance=24pt]
    	\coordinate
    			child{ edge from parent [snake edge]
    				child { edge from parent [snake edge] }
    				child { edge from parent [snake edge] }			
    			} ;	
    \end{tikzpicture}
    ~ + 3 ~
    \begin{tikzpicture}[scale=0.3,thick]	
    	\tikzstyle{level 1}=[level distance=24pt]
    	\coordinate
    			child{ edge from parent [snake edge]
    				child { edge from parent [snake edge]
    					child { edge from parent [snake edge] }
    					child { edge from parent [snake edge] }	
    				}
    				child { edge from parent [snake edge] }			
    			} ;	
    \end{tikzpicture}
    ~ + 3 ~
    \begin{tikzpicture}[scale=0.3,thick]	
    	\tikzstyle{level 1}=[level distance=24pt]
    	\coordinate
    			child{ edge from parent [snake edge]
    				child { edge from parent [snake edge] }
    				child { edge from parent [snake edge]
    					child { edge from parent [snake edge] }
    					child { edge from parent [snake edge] }	
    				}			
    			} ;	
    \end{tikzpicture}
    ~ + ~ 4 ~
    \begin{tikzpicture}[scale=0.3,thick]	
    	\tikzstyle{level 1}=[level distance=24pt, sibling distance=24pt]
    	\tikzstyle{level 2}=[level distance=24pt, sibling distance=24pt]
    	\tikzstyle{level 3}=[level distance=24pt, sibling distance=20pt]
    	\coordinate
    			child{ edge from parent [snake edge]
    				child { edge from parent [snake edge]
    					child { edge from parent [snake edge] }
    					child { edge from parent [snake edge] }
    				}
    				child { edge from parent [snake edge]
    					child { edge from parent [snake edge] }
    					child { edge from parent [snake edge] }	
    				}			
    			} ;	
    \end{tikzpicture}
    + ~ \dots \right) 
    \times ~
    \begin{tikzpicture}[scale=0.3,thick]	
    	\tikzstyle{level 1}=[level distance=24pt]
    	\coordinate
    			child{ edge from parent [snake edge]
    				child { edge from parent [full edge] }
    				child { edge from parent [double edge] }			
    			} ;	
    \end{tikzpicture}
    %%%new line%%%%%%%%
    \nonumber \\ & 
    ~~~ \times \left(~
    \begin{tikzpicture}[scale=0.3,thick]	
    	\tikzstyle{level 1}=[level distance=24pt]
    	\coordinate
    			child{ edge from parent [full edge] 	} ;	
    \end{tikzpicture}
    ~ + ~
    \begin{tikzpicture}[scale=0.3,thick]	
    	\tikzstyle{level 1}=[level distance=24pt]
    	\coordinate
    			child{ edge from parent [full edge]
    				child { edge from parent [full edge] }
    				child { edge from parent [dash edge] }			
    			} ;	
    \end{tikzpicture}
    ~ + ~
    \begin{tikzpicture}[scale=0.3,thick]	
    	\tikzstyle{level 1}=[level distance=24pt]
    	\coordinate
    			child{ edge from parent [full edge]
    				child { edge from parent [dot edge] }
    				child { edge from parent [full edge] }			
    			} ;	
    \end{tikzpicture}
    ~ + ~
    \begin{tikzpicture}[scale=0.3,thick]	
    	\tikzstyle{level 1}=[level distance=24pt]
    	\coordinate
    			child{ edge from parent [full edge]
    				child { edge from parent [full edge]
    					child { edge from parent [full edge] }	
    					child { edge from parent [dash edge] }	
    				}
    				child { edge from parent [dash edge] }			
    			} ;	
    \end{tikzpicture}
    ~ + ~
    \begin{tikzpicture}[scale=0.3,thick]	
    	\tikzstyle{level 1}=[level distance=24pt]
    	\coordinate
    			child{ edge from parent [full edge]
    				child { edge from parent [full edge]
    					child { edge from parent [dot edge] }	
    					child { edge from parent [full edge] }	
    				}
    				child { edge from parent [dash edge] }			
    			} ;	
    \end{tikzpicture}
    ~ + ~
    \begin{tikzpicture}[scale=0.3,thick]	
    	\tikzstyle{level 1}=[level distance=24pt]
    	\coordinate
    			child{ edge from parent [full edge]
    				child { edge from parent [dot edge]
    					child { edge from parent [dot edge] }	
    					child { edge from parent [dot edge] }	
    				}
    				child { edge from parent [full edge] }			
    			} ;	
    \end{tikzpicture}
    ~ + ~ \dots
    \right)
    %%%new line%%%%%%%%
    \nonumber \\ &~~~
    \times \left( ~
    \begin{tikzpicture}[scale=0.3,thick]	
    	\tikzstyle{level 1}=[level distance=24pt]
    	\coordinate
    			child{ edge from parent [double edge] 	} ;	
    \end{tikzpicture}
    ~ + ~
    \begin{tikzpicture}[scale=0.3,thick]	
    	\tikzstyle{level 1}=[level distance=24pt]
    	\coordinate
    			child{ edge from parent [double edge]
    				child { edge from parent [double edge] }
    				child { edge from parent [dot edge] }			
    			} ;	
    \end{tikzpicture}
    ~ + ~
    \begin{tikzpicture}[scale=0.3,thick]	
    	\tikzstyle{level 1}=[level distance=24pt]
    	\coordinate
    			child{ edge from parent [double edge]
    				child { edge from parent [dash edge] }
    				child { edge from parent [double edge] }			
    			} ;	
    \end{tikzpicture}
    ~ + ~
    \begin{tikzpicture}[scale=0.3,thick]	
    	\tikzstyle{level 1}=[level distance=24pt]
    	\coordinate
    			child{ edge from parent [double edge]
    				child { edge from parent [double edge]
    					child { edge from parent [double edge] }	
    					child { edge from parent [dot edge] }	
    				}
    				child { edge from parent [dot edge] }			
    			} ;	
    \end{tikzpicture}
    ~ + ~
    \begin{tikzpicture}[scale=0.3,thick]	
    	\tikzstyle{level 1}=[level distance=24pt]
    	\coordinate
    			child{ edge from parent [double edge]
    				child { edge from parent [double edge]
    					child { edge from parent [dash edge] }	
    					child { edge from parent [double edge] }	
    				}
    				child { edge from parent [dot edge] }			
    			} ;	
    \end{tikzpicture}
    ~ + ~
    \begin{tikzpicture}[scale=0.3,thick]	
    	\tikzstyle{level 1}=[level distance=24pt]
    	\coordinate
    			child{ edge from parent [double edge]
    				child { edge from parent [dash edge]
    					child { edge from parent [dash edge] }	
    					child { edge from parent [dash edge] }	
    				}
    				child { edge from parent [double edge] }			
    			} ;	
    \end{tikzpicture}
    ~ + ~ \dots
     \right) ~.
\end{align}
The latter two series in parentheses we recognise as the generating functions \eqref{eq-genRootPath} of rooted paths on the left and right arcs, respectively. 
So let us write
\begin{equation}
    \begin{tikzpicture} [thick, scale=0.5]	
    	\tikzstyle{level 1}=[level distance=24pt]
    	\coordinate
    			child{ edge from parent [redressed edge] } ;	
    \end{tikzpicture}
    ={\gJ}(\xi,x,z,\zeta) \quad , \qquad\qquad
    \begin{tikzpicture} [thick, scale=0.5]	
    	\tikzstyle{level 1}=[level distance=24pt]
    	\coordinate
    			child{ edge from parent [redressed double edge] } ;	
    \end{tikzpicture}
    ={\gJ}(\xi,x,\bar{z},\bar{\zeta}) ~,
\end{equation}
using which we have
\begin{equation}
    \begin{tikzpicture}[scale=0.3,thick]	
    	\tikzstyle{level 1}=[level distance=24pt]
    	\coordinate
    			child{ edge from parent [snake edge] [fill=black] circle (8pt)
    				child { edge from parent [full edge] }
    				child { edge from parent [double edge] }			
    			} ;	
    \end{tikzpicture}
    ~=
    \left(
    \begin{tikzpicture}[scale=0.3,thick]	
    	\tikzstyle{level 1}=[level distance=24pt]
    	\coordinate
    			child{ edge from parent [snake edge] } ;	
    \end{tikzpicture}
    ~ + 2 ~
    \begin{tikzpicture}[scale=0.3,thick]	
    	\tikzstyle{level 1}=[level distance=24pt]
    	\coordinate
    			child{ edge from parent [snake edge]
    				child { edge from parent [snake edge] }
    				child { edge from parent [snake edge] }			
    			} ;	
    \end{tikzpicture}
    ~ + 3 ~
    \begin{tikzpicture}[scale=0.3,thick]	
    	\tikzstyle{level 1}=[level distance=24pt]
    	\coordinate
    			child{ edge from parent [snake edge]
    				child { edge from parent [snake edge]
    					child { edge from parent [snake edge] }
    					child { edge from parent [snake edge] }	
    				}
    				child { edge from parent [snake edge] }			
    			} ;	
    \end{tikzpicture}
    ~ + 3 ~
    \begin{tikzpicture}[scale=0.3,thick]	
    	\tikzstyle{level 1}=[level distance=24pt]
    	\coordinate
    			child{ edge from parent [snake edge]
    				child { edge from parent [snake edge] }
    				child { edge from parent [snake edge]
    					child { edge from parent [snake edge] }
    					child { edge from parent [snake edge] }	
    				}			
    			} ;	
    \end{tikzpicture}
    ~ + 4 ~
    \begin{tikzpicture}[scale=0.3,thick]	
    	\tikzstyle{level 1}=[level distance=24pt, sibling distance=24pt]
    	\tikzstyle{level 2}=[level distance=24pt, sibling distance=24pt]
    	\tikzstyle{level 3}=[level distance=24pt, sibling distance=20pt]
    	\coordinate
    			child{ edge from parent [snake edge]
    				child { edge from parent [snake edge]
    					child { edge from parent [snake edge] }
    					child { edge from parent [snake edge] }
    				}
    				child { edge from parent [snake edge]
    					child { edge from parent [snake edge] }
    					child { edge from parent [snake edge] }	
    				}			
    			} ;	
    \end{tikzpicture}
     + \dots
    \right)
    \times~
    \begin{tikzpicture}[scale=0.3,thick]	
    	\tikzstyle{level 1}=[level distance=24pt]
    	\coordinate
    			child{ edge from parent [snake edge]
    				child { edge from parent [redressed edge] }
    				child { edge from parent [redressed double edge] }			
    			} ;	
    \end{tikzpicture} ~.
\end{equation}
The remaining series left to be summed can be seen to be $\sum_n (n+1)\vC_n\lambda^n$ which we can write as $\pdd{}{\lambda}\lambda {\gC}(\lambda)$. 
Altogether we find for the generating function
\begin{equation}
    \mu {\gM}(\xi,x,z,\zeta,\bar{z},\bar{\zeta},\lambda) = \mu {\gJ}(\xi,x,z,\zeta) \times \gJ(x,\xi,\bar{z},\bar{\zeta}) \times \pdd{}{\lambda}  \left[ \lambda {\gC}(\lambda) \right] ~.
\end{equation}
The derivative of ${\gC}(\lambda)$ can be easily determined given \eqref{eq-Cquad}, which yields
%\begin{align}
$\pdd{}{\lambda}{\gC}(\lambda) = {\gC}(\lambda)^2 + 2\lambda {\gC}(\lambda) \pdd{}{\lambda} {\gC}(\lambda) = \frac{ {\gC}(\lambda)^2 }{ 1 - 2\lambda {\gC}(\lambda) }$ and
$\pdd{}{\lambda}  \left[ \lambda {\gC}(\lambda) \right] = {\gC}(\lambda) + \frac{ \lambda {\gC}(\lambda)^2 }{ 1 - 2\lambda {\gC}(\lambda) }  = \frac{ {\gC}(\lambda) -  \lambda {\gC}(\lambda)^2 }{ 1 - 2\lambda {\gC}(\lambda) }$.
%\end{align}
Finally we arrive at the generating function
\begin{equation}
    {\gM}(\xi,x,z,\zeta,\bar{z},\bar{\zeta},\lambda) = \frac{1}{1 - z {\gC}(x) - \zeta {\gC}(\xi)} \: \frac{1}{1 - \bar{z} {\gC}(x) - \bar{\zeta} {\gC}(\xi)} \: \frac{1}{ 1 - 2\lambda {\gC}(\lambda) } ~.
\end{equation}
The separation between the two leaves is given by the number of vertices ($x$) sitting between the two arcs of the path plus the number of left-turning vertices ($z$) in the left arc of the path plus the number of right-turning vertices ($\bar{z}$) in the right arc of the path. 
With this in mind we attach a dummy variable to these variables and look for coefficients of $y^{s}$, giving
%we can write the generating function for leaf to leaf paths with a fixed separation $s$ as
\begin{equation}
    \vM^{(s)}(\xi,x,z,\zeta,\bar{z},\bar{\zeta},\lambda) = [y^s] {\gM}(\xi,y x,y z,\zeta,y \bar{z},\bar{\zeta},\lambda,\mu) ~.
\end{equation}
To find the average length of a path between leafs with separation $s$ we first need to sum over all trees with $n$ vertices, the number of paths with length $k+l+\bar{k}+\bar{l}+1$ multiplied by that length. 
The factor of $1$ here comes in because we always need to account for the initial bifurcation given by the $\mu$ vertex. 
We therefore define
\begin{equation}
    \vP^{(s)}_n = \sum_{i+j+k+l+\bar{k}+\bar{l}+m+1=n} (k+l+\bar{k}+\bar{l}+1)\vM^{(s)}_{ijkl\bar{k}\bar{l}m} ~.
    \label{eq-vPn}
\end{equation}
We proceed to manipulate this expression in a similar manner to Eq.\ \eqref{eq-vEn}. 
First using \eqref{eq-genfuneuler} we simplify the summand
\begin{align}
    &(k+l+\bar{k}+\bar{l}+1)\vM^{(s)}_{ijkl\bar{k}\bar{l}m} \nonumber \\
    &= (k+l+\bar{k}+\bar{l}+1) [\xi^i x^j z^k \zeta^l \bar{z}^{\bar{k}} \bar{\zeta}^{\bar{l}} \lambda^m y^s]  \left\{ {\gM}(\xi,yx,yz,\zeta,y\bar{z},\bar{\zeta} ,\lambda) \right\}  \nonumber \\
    &=  [\xi^i x^j z^k \zeta^l \bar{z}^{\bar{k}} \bar{\zeta}^{\bar{l}} \lambda^m y^s]  \left\{ \left( z\pdd{}{z} + \bar{z} \pdd{}{\bar{z}} + \zeta \pdd{}{\zeta} + \bar{\zeta}\pdd{}{\bar{\zeta}} +1 \right) {\gM}(\xi,yx,yz,\zeta,y\bar{z},\bar{\zeta} ,\lambda) \right\}
\end{align}
to which the derivatives can be performed straightforwardly,
\begin{align}
    &\left( z\pdd{}{z} + \bar{z} \pdd{}{\bar{z}} + \zeta \pdd{}{\zeta} + \bar{\zeta}\pdd{}{\bar{\zeta}} +1 \right) {\gM}(\xi,yx,yz,\zeta,y\bar{z},\bar{\zeta},\lambda) \nonumber\\
    & \qquad = \frac{ 1 }{ 1 - 2\lambda {\gC}(\lambda) }
    \: \frac{1}{ 1 - yz {\gC}(yx) - \zeta {\gC}(\xi) }
    \: \frac{1}{ 1 - y\bar{z} {\gC}(yx) - \bar{\zeta} {\gC}(\xi)} \nonumber \\
    & \qquad \qquad \times \left[ \frac{yz {\gC}(yx) + \zeta {\gC}(\xi)}{ 1 - yz {\gC}(yx) - \zeta {\gC}(\xi) } + \frac{ y\bar{z} {\gC}(yx) + \bar{\zeta} {\gC}(\xi) }{ 1 - y\bar{z} {\gC}(yx) - \bar{\zeta} {\gC}(\xi)}  +1 \right] .
\end{align}
Applying the formula \eqref{eq-sumout} to $\xi$, $x$, $z$, $\bar{z}$, $\zeta$ , $\bar{\zeta}$ and $\lambda$, $\vP^{(s)}_n$ is reduced to
\begin{align}
    \vP^{(s)}_n &= [x^{n-1} y^s] \left\{ \frac{ 1 }{ 1 - 2 x {\gC}(x) } \left[ \frac{ 2[ x {\gC}(x) + xy {\gC}(xy) ]  }{ [ 1 -x {\gC}(x) - xy {\gC}(xy) ]^3 } + \frac{ 1 }{ [ 1 -x {\gC}(x) - xy {\gC}(xy) ]^2 }\right]  \right\} ~.
\end{align}
Simplifying slightly using \eqref{eq-genfunlower} we arrive at
\begin{equation}
    \vP^{(s)}_n = [x^n y^s] \left\{
    \frac{ 1 }{ 1 - 2 x {\gC}(x) }
    \frac{ 1 }{ [ 1 -x {\gC}(x) - xy {\gC}(xy) ]^2 } \left[  \frac{ 2 }{ 1 -x {\gC}(x) - xy {\gC}(xy)  }  - 1 \right]  \right\} ~.
    \label{eq-vP}
\end{equation}
This summed length with separation $s$ on a tree with $n$ vertices can be further simplified as follows
\begin{align}
    \vP^{(s)}_n &= [x^{n-1} y^s] \left\{
    \left[ \frac{2}{[1-x{\gC}(x)-xy{\gC}(xy)]^3}- \frac{1}{[1-x{\gC}(x)-xy{\gC}(xy)]^2} \right]  \pdd{x{\gC}(x)}{x}  \right\} \nonumber\\
    &= [x^{n-1-s} z^s] \left\{
    \left[ \frac{2}{[1-x{\gC}(x)-z{\gC}(z)]^3} - \frac{1}{[1-x{\gC}(x)-z{\gC}(z)]^2} \right]  \pdd{x{\gC}(x)}{x}  \right\} \nonumber\\
    &= [x^{n-1-s} z^s] \left\{ \pdd{}{x}
    \left[ \frac{1}{[1-x{\gC}(x)-z{\gC}(z)]^2} - \frac{1}{[1-x{\gC}(x)-z{\gC}(z)]} \right]  \right\} \nonumber\\
    &= (n-s) [x^{n} y^s] \left\{
    \frac{1}{[1-x{\gC}(x)-xy{\gC}(xy)]^2} - \frac{1}{[1-x{\gC}(x)-xy{\gC}(xy)]} \right\}  
    %\nonumber\\ &=(n-s) \vL^{(s)}_{n} 
    ~,
\end{align}
where we have used $z = xy$.
By comparison with Eq.\ (\ref{eq-depthGF}) we conclude that the summed length of leaf to leaf paths is related directly to the sum of the depth of the leaves by
\begin{equation}
    \vP^{(s)}_n = (n-s) \vL^{(s)}_{n}
    \label{eq-PnsL}
\end{equation}
and hence the \emph{average} length of a leaf to leaf path with separation $s$ on a tree with $n$ vertices is
\begin{equation}
    \aP^{(s)}_n = \frac{(n-s) \vL^{(s)}_{n}}{(n-s) \vC_{n}} = \frac{\vL^{(s)}_{n}}{\vC_{n}} = d^{(s)}_{n} ~.
    \label{eq-AnsLC}
\end{equation}
We therefore see that the path length when averaged over both the set of all trees as well as the all pairs of leaves with separation $s$ within each tree is identical to the average depth of a rooted path with penetration $s$.
The explicit form of which is given by Eq.\ (\ref{eq-av_depth}).

The asymptotic properties of the path length in the limit of large $n$ can be found using Stirling's approximation
\begin{equation}
n! \sim \sqrt{2\pi n}\left( \frac{n}{e} \right)^n ~.
\end{equation}
Accordingly, we find that
$%\begin{equation}
\vC_n \sim \frac{1}{\sqrt{\pi}} \frac{1}{n\sqrt{n}} 2^{2n}
$ %\end{equation}
for large $n$.
Therefore, we conclude from Eq.\ \eqref{eq-av_depth} that
\begin{displaymath}
\aP_\infty^{(s)} \equiv \lim_{n \to \infty} \aP_{n}^{(s)} = \frac{(s+1)(2s+1)}{4^{s-1}} \vC_s -1 ~.
\end{displaymath}
In the continuum limit, $0 \ll s \ll n$, we have
\begin{equation}
\aP_{\infty}^{(s)} \sim \sqrt{\frac{64 s}{\pi}} ~.
\end{equation}

\begin{figure}[ht]
    \centering
    \includegraphics[width=0.75\columnwidth]{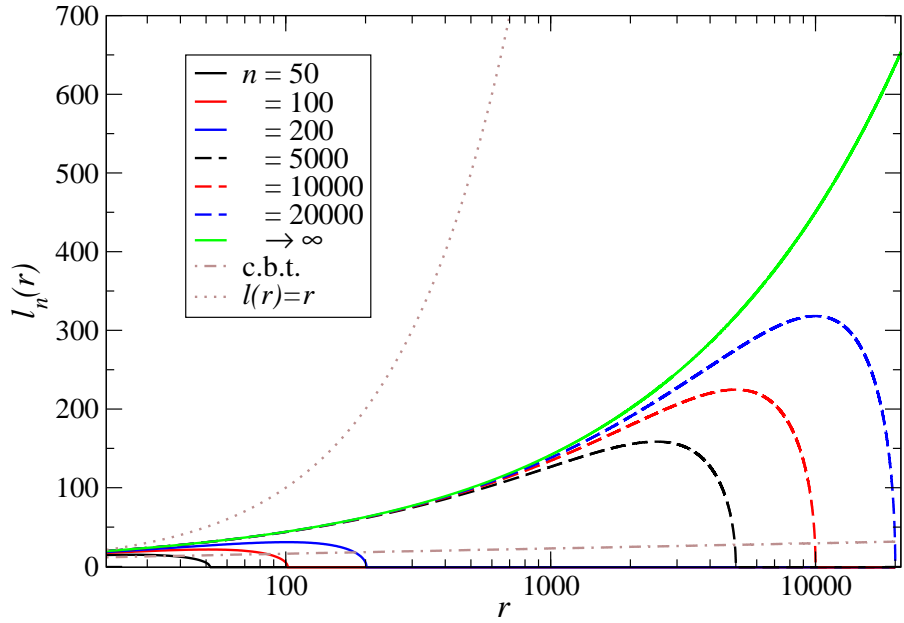}
    \caption{Average length of a leaf to leaf path $\aP_n(r)$ versus separation $r$ for small $n= 50$ (solid black line), $100$ (solid red), $200$ (solid blue) and large $n= 5,000$ (dashed black line), $10,000$ (dashed red), $20,000$ (dashed blue). 
      The solid green line denotes $\aP_{\infty}(r)$, the dotted line corresponds to $r$ while the dashed-dotted line shows the corresponding result for complete binary trees.\cite{GolRR15}
    \label{fig-AFinal}}
\end{figure}

In Fig.\ \ref{fig-AFinal} we transform back into an expression for $r=s+1$, $\aP_{n}(r)$, to compare with previous results for the complete binary tree \cite{GolRR15}, shown as a dashed line.
We show the average path length for various separations as well as the large $n$ limit.
This clearly illustrates that the different tree geometry can dramatically change the scaling of the path length.
%%%%%%%%%%%%%%%%%%%%%%%%%%%%%%%%%%%%%%%%
%%%%%%%%%%%%%%%%%%%%%%%%%%%%%%%%%%%%%%%%

\begin{figure}[ht]
    \centering
    \includegraphics[width=0.9\textwidth]{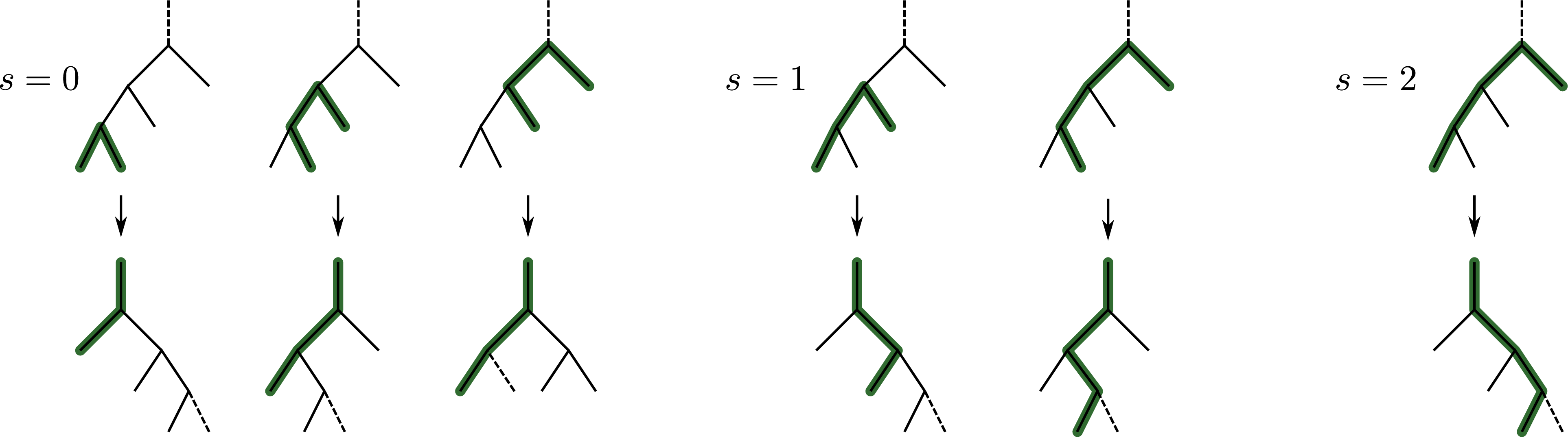}
    \caption{Graphical representation of the equivalence between path length on a Catalan graph with $n=3$ nodes and a rooted path connecting the same nodes created by deforming the tree. 
      All possible paths on a particular tree are shown, paths with separation $s$ are mapped to trees with penetration $s$. Symbols and lines are as in Fig.\ \ref{fig-schematic}.
  \label{fig-geom_proof_leg}}
\end{figure}

\begin{figure}[ht]
    \centering
    \includegraphics[width=0.5\columnwidth]{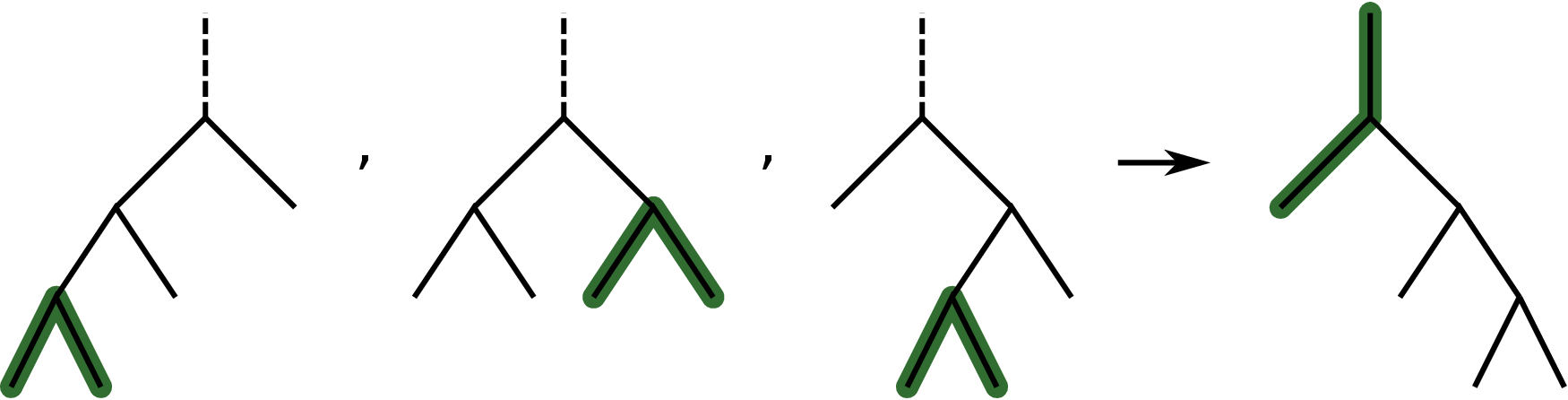}
    \caption{Example of the degeneracy of rooted paths, showing the $(n-s)=3$ cases where $n=3$ and $s=0$ that map to the same rooted tree. 
      Symbols and lines are as in Fig.\ \ref{fig-schematic}.
    \label{fig-geom_proof_degen}}
\end{figure}

The connection found in \eqref{eq-PnsL} can be heuristically understood as follows:
Take a leaf-to-leaf path and deform the tree such that the left hand leaf is above the root, making the first vertex the new root. 
Now it is clear that any path with separation $s$ can be expressed as a leg depth of penetration $s$, as shown in Fig.\ \ref{fig-geom_proof_leg}. 
As the set of Catalan trees is \emph{complete} in the sense that all possible binary trees are part of the set of Catalan trees, the newly deformed tree is also one of the set.
%%%%%%%%%%%%%%%%%%%%%%%%%%%%%%%%%%%%%%%%
%%%%%%%%%%%%%%%%%%%%%%%%%%%%%%%%%%%%%%%%
By observation, in for example Fig.\ \ref{fig-geom_proof_degen}, one can see that $(n-s)$ paths map to each unique rooted tree depth. 
The degeneracy coming from the number of possible vertices to the right of the right hand leaf, which can be the original root, is highlighted in Figures \ref{fig-geom_proof_leg} and \ref{fig-geom_proof_degen} by the dashed edge.
%%%%%%%%%%%%%%%%%%%%%%%%%%%%%%%%%%%%%%%%
%%%%%%%%%%%%%%%%%%%%%%%%%%%%%%%%%%%%%%%%

%%%%%%%%%%%%%%%%%%%%%%%%%%%%%%%%%%%%%%%%%%%%%%%%%%%%%%%%%%%%%%%%%%%
\section{Conclusions}
%%%%%%%%%%%%%%%%%%%%%%%%%%%%%%%%%%%%%%%%%%%%%%%%%%%%%%%%%%%%%%%%%%%
As shown in Eq.\ \eqref{eq-AnsLC}, within Catalan trees the average path length between leaves with separation $s$ is equal to the average depth of a leaf with penetration $s$.
In Figures \ref{fig-geom_proof_leg} and \ref{fig-geom_proof_degen}, we show that there is an intriguing geometric interpretation of this result.
After completion of our work, we were made aware of a proof of the average depth of a leaf of penetration $p$ (Eq.\ (\ref{eq-av_depth})) by Kirchenhofer \cite{Kir83} and a later generalisation to all statistical moments.\cite{PanP02}
As the generating functions of the leaf depths and path lengths are directly related, the results for the higher statistical moments should also be valid here.

The final results for $\aP_{n}(r)$ and $\aP_{\infty}(r)$ are quite surprising to us when comparing, as done in Fig.\ \ref{fig-AFinal}, with the corresponding expressions ${\cal L}_n(r)$ and ${\cal L}_{\infty}(r)$ for complete binary trees \cite{GolRR15} and the tree graphs emerging from the tree tensor network\cite{GolR14}. 
We find that the $\aP$-path lengths take larger values than the ${\cal L}$'s. 
This can be rationalized as follows: in order for $n$ vertices to form into a complete binary graph, those vertices are ``packed'' very densely, hence giving rise to a small ${\cal L}_n(r)$. 
On the other hand, Catalan graphs with $n$ vertices can have a much more extended structure, leading to larger path lengths $\aP_n(r)$ for the same $r$.
Comparing to the tree tensor networks, this result shows that our naive expectation expressed in the introduction, is wrong: the path lengths are quite different for larger $r$ values and hence the implied correlation functions have a very different long distance behaviour. 
Just as for complete binary graphs, it remains to be seen what Hamiltonians, if any, correspond to these correlations.

%%%%%%%%%%%%%%%%%%%%%%%%%%%%%%%%%%%%%%%%%%%%%%%%%%%%%%%%%%%%%%%%%%%
\section*{Acknowledgments}
%%%%%%%%%%%%%%%%%%%%%%%%%%%%%%%%%%%%%%%%%%%%%%%%%%%%%%%%%%%%%%%%%%%
We thank the EPSRC for financial support (EP/J003476/1) and provision of computing resources through the MidPlus Regional HPC Centre (EP/K000128/1). 
We are grateful to H.\ Prodinger for pointing out relevant references \cite{Kir83,Kir83_2,PanP02}. No new data were created during this study.

%
%%%%%%%%%%%%%%%%%%%%%%%%%%%%%%%%%%%%%%%%%%%%%%%%%%%%%%%%%%%%%%%%%%%%
%\section{Author contributions statement}
%%%%%%%%%%%%%%%%%%%%%%%%%%%%%%%%%%%%%%%%%%%%%%%%%%%%%%%%%%%%%%%%%%%%
%A.M.G.\ proposed the problem, performed analysis and prepared the article for publication.
%J.M.F.\ found the successful method for relating depth and path length, and created a first draft of the solution.
%S.A.R., M.B.\ and G.R.\ provided mathematical insight and, along with A.M.G., calculated the average leaf depth in the absence of Kirchenhoffer's derivation. 
%R.A.R.\ oversaw the project, provided the proof for the \emph{incomplete Segner relation} with J.M.F., and wrote the first full paper draft. 
%%%%%%%%%%%%%%%%%%%%%%%%%%%%%%%%%%%%%%%%%%%%%%%%%%%%%%%%%%%%%%%%%%%%
%\section{Additional information}
%%%%%%%%%%%%%%%%%%%%%%%%%%%%%%%%%%%%%%%%%%%%%%%%%%%%%%%%%%%%%%%%%%%%
%\textbf{Competing financial interests} The authors declare no competing financial interests. 

\numberwithin{equation}{section}
\renewcommand{\thesection}{\Alph{section}}
\renewcommand{\theequation}{\Alph{section}.\arabic{equation}}
\setcounter{equation}{0}
\setcounter{section}{0}

%%%%%%%%%%%%%%%%%%%%%%%%%%%%%%%%%%%%%%%%%%%%%%%%%%%%%%%%%%%%%%%%%%%
\section{Appendix}
%%%%%%%%%%%%%%%%%%%%%%%%%%%%%%%%%%%%%%%%%%%%%%%%%%%%%%%%%%%%%%%%%%%

%%%%%%%%%%%%%%%%%%%%%%%%%%%%%%%%%%%%%%%%%%%%%%%
\subsection{More on generating functions}
\label{sec:generating_functions}
%%%%%%%%%%%%%%%%%%%%%%%%%%%%%%%%%%%%%%%%%%%%%%%
We make extensive use of generating functions \cite{Kos09,Wil94} as they offer a convenient way to manipulate relations between the set of Catalan numbers. 
More generally, they serve as a very useful formal device for manipulating a series and for determining the properties of unknown series \cite{Wil94}.
Given a general (infinite) series $a_0,a_1,a_2,a_3\dots$, it is always possible to formally define a function
\begin{equation}
\label{eq-defgenfun}
    a(x) = a_0 + a_1 x + a_2 x^2 + a_3 x^3 + \dots = \sum_{n=0}^{\infty} a_n x^n ~,
\end{equation}
which we call the \emph{generating function} of the sequence $a_n$.
We can also invert the above definition and instead say that $a_n$ is the coefficient of $x^n$ in the Taylor series representation of $a(x)$ about zero. 
We employ the notation\cite{Wil94} $a_n = [x^n] \{ a(x) \}$.
For example, squaring \eqref{eq-Csum} gives,
\begin{align}
    \gC^{2}(x) &= \vC_{0}^{2} + (\vC_{0}\vC_{1} + \vC_{1}\vC_{0})x + \dots + (\vC_{0}\vC_{n} + \vC_{1}\vC_{n-1} + \dots + \vC_{n}\vC_{0})x^{n} + \dots \nonumber \\
               &= \vC_{1} + \vC_{2}x + \vC_{3}x^{2} + \dots + \vC_{n+1}x^{n} + \dots
               = \frac{\gC(x)-\vC_{0}}{x} ~,
    \label{eq-Cderiv}
\end{align}
and we have used the final part of Eq.\ \eqref{eq-Catalan_recursion} in each term.
Multiplication by $x$ results in \eqref{eq-Cquad}.

Below, we list some particular generating functions and their associated series coefficients. The series coefficients of $a(x)$ can be expressed as \cite{Wil94}
\begin{equation}
    [x^n]\{a(x)\} = \left. \frac{1}{n!}\pdd{^na(x)}{n^n} \right\vert_{x=0}
    = \oint \frac{dx}{2\pi i}  \frac{1}{x} \frac{a(x)}{x^n} ~.
    \label{eq-defgenfunascontour}
\end{equation}
A useful identity is the \emph{Lagrange inversion formula} \cite{Wil94,SedF13}: Suppose $u$ is a function which can be expressed as $u(x)=x\phi(u)$ where the function $\phi$ satisfies $\phi(0)=1$, then for a given function $f$, we have
\begin{equation}
    [x^n]\{f(u(x))\} = \frac{1}{n}[u^{n-1}]\{f'(u)\phi(u)^n\} ~.
    \label{eq-lagrange}
\end{equation}
Generating functions can also be defined for multi-indexed series $a_{ijk\dots}$ \cite{Wil94}. We will extend our notation so that \\$[x_1^{n_1}x_2^{n_2}\dots x_r^{n_r}]\{f(x_1,x_2,\dots,x_r)\}$ is the coefficient of $x_1^{n_1}x_2^{n_2}\dots x_r^{n_r}$ in the Taylor series of $f(x_1,x_2,\dots,x_r)$.
A convenient identity for dealing with generating functions of many variables is
\begin{equation}
    \sum_{r=0}^{n} [x^r y^{n-r}]\{a(x,y)\} = [x^n]\{ a(x,x) \} ~.
    \label{eq-sumout}
\end{equation}

As an example of the use of the notation for generating functions,\cite{Wil94} we list some particular generating functions and their associated series coefficients,
\begin{equation}
    %\label{eq-specgenfunpow}
    %& 
    [x^n]\{ x^p \} = \delta_{n,p} ~, \quad%\\
    %\label{eq-specgenfunexp}
    %& 
    [x^n]\{ e^x \} = 1/n! ~, \quad%\\
    %\label{eq-specgenfungeo}
    %& 
    [x^n]\{ \tfrac{1}{1-x} \} = 1 ~, \quad%\\
    %\label{eq-specgenfunbinom}
    %& 
    [x^n]\{ (1+x)^p \} = \binom{p}{n} ~.
\end{equation}
If $[x^n]\{a(x)\}=a_n$ and $[x^n]\{b(x)\}=b_n$, one finds
\begin{equation}
    [x^n]\{ a(x)b(x) \} = \sum_{k=0}^n a_k b_{n-k} ~.
    \label{eq-genfunprod}
\end{equation}
Raising and lowering of indices in series coefficients can be achieved using 
\begin{align}
    \label{eq-genfunraise}
    [x^{n-1}]\{ a(x) \} &= [x^n]\{ x a(x) \} ~,\\
    \label{eq-genfunlower}
    [x^{n+1}]\{ a(x) \} &=[x^n]\{ \tfrac{ a(x)-a(0) }{ x } \} ~.
\end{align}
Differentiating Eq.\ \eqref{eq-defgenfun} we notice that
\begin{equation}
    [x^n]\left\{ \pdd{a(x)}{x} \right\} = n[x^{n+1}]\{ a(x) \} ~,
    \label{eq-genfunderiv}
\end{equation}
combining this with Eq.\ \eqref{eq-genfunraise} we get the useful relation
\begin{equation}
    n[x^n]\{ a(x) \} = [x^n]\left\{ x \pdd{}{x} a(x)\right\} ~.
    \label{eq-genfuneuler}
\end{equation}
From the series definition it should be clear that constant multiples of the argument can just be brought out front as follows
\begin{equation}
    [x^n]\{a(cx)\} = c^n[x^n]\{a(x)\} ~.
    \label{eq-genfunargmult}
\end{equation}
%
%The proof of the Lagrange inversion formula \eqref{eq-lagrange} is very simple, applying the identities stated above \cite{Wil94}
%\begin{align*}
%    [x^n]\{f(u(x))\} &= [x^n]\left\{ \frac{1}{n} x\pdd{}{x} f(u(x))\right\} \\
%    &=\frac{1}{n} \oint \frac{dx}{2\pi i} \frac{1}{x} \frac{1}{x^n}~  x \pdd{f}{u} \dd{u}{x} \\
%    &= \frac{1}{n} \oint \frac{dx}{2\pi i } \left( \frac{\phi(u)}{u}\right)^n \pdd{f}{u} \dd{u}{x} \\
%    &= \frac{1}{n} \oint \frac{dx}{2\pi i u} \dd{u}{x} \frac{\phi(u)}{u^{n-1}} f'(u) \\
%    &= \frac{1}{n} \oint \frac{du}{2\pi i u}  \frac{\phi(u)}{u^{n-1}} f'(u) \\
%    &=\frac{1}{n}[u^{n-1}]\{f'(u)\phi(u)^n\} ~.
%\end{align*}
%%
%Similarly, \eqref{eq-sumout} can be shown as follows,
%\begin{align*}
%    \sum_{r=0}^{n} [x^r y^{n-r}]\{a(x,y)\}  &= \left. \sum_{r=0}^n \frac{1}{(n-r)!} \pdd{^{n-r}}{x^{n-r}}\frac{1}{r!} \pdd{^{r}}{x^{r}} a(x,y)\right\vert_{x,y=0} \\
%    &= \frac{1}{n!} \left. \sum_{r=0}^n \binom{n}{r}\pdd{^{n-r}}{x^{n-r}} \pdd{^{r}}{x^{r}} a(x,y)\right\vert_{x,y=0} \\
%    &=\left. \frac{1}{n!} \left( \pdd{}{x} + \pdd{}{y} \right)^n a(x,y)\right\vert_{x,y=0} \\
%    &= \left. \frac{1}{n!} \pdd{^n}{t^n}a(t+s,t-s)\right\vert_{s,t=0} \\
%    &= \left. \frac{1}{n!} \pdd{^n}{t^n}a(t,t)\right\vert_{t=0} \\
%    &= [t^n]\{ a(t,t) \} ~,
%\end{align*}
%where we have used $t=\tfrac{1}{2}(x+y)$ and $s=\tfrac{1}{2}(x-y)$.
%We can rename $t$ to $x$ as it is a dummy variable.
For Eq.\ \eqref{eq-vEn} it is useful to notice that, using \eqref{eq-Cquad},
\begin{equation}
    {\gK}(x,x)={\gC}(x) ~, \quad
    \pdd{{\gK}(x,z)}{z}={\gC}(x) {\gK}(x,z)^2 ~.
    %\label{eq-dkdz}
    \label{eq-kxx}
\end{equation}
%and
%\begin{equation}
%    \pdd{{\gK}(x,z)}{z}={\gC}(x) {\gK}(x,z)^2~.
%    \label{eq-dkdz}
%\end{equation}

%%%%%%%%%%%%%%%%%%%%%%%%%%%%%%%%%%%%%%%%%%%%%%%
\subsection{Derivation of the average depth of a leaf}
\label{sec:DepthDeriv}
%%%%%%%%%%%%%%%%%%%%%%%%%%%%%%%%%%%%%%%%%%%%%%%
Equation (\ref{eq-depthGF}) can be simplified by multiplying top and bottom by $x{\gC}(x)-xy{\gC}(xy)$, due to $1/[ 1-x{\gC}(x)-xy {\gC}(xy)] = [ x{\gC}(x)-xy{\gC}(xy) ]/( x-xy )$.
% as follows
% \begin{align}
% \frac{1}{ 1-x{\gC}(x)-xy {\gC}(xy)} &=\frac{1}{ 1-x{\gC}(x)-xy {\gC}(xy)} \frac{ x{\gC}(x)-xy{\gC}(xy) }{ x{\gC}(x)-xy{\gC}(xy) } \nonumber \\
% &=\frac{ x{\gC}(x)-xy{\gC}(xy) }{ x{\gC}(x)-xy{\gC}(xy) -x^2{\gC}(x)^2+x^2y^2{\gC}(xy)^2} \nonumber \\
% &=\frac{ x{\gC}(x)-xy{\gC}(xy) }{ x({\gC}(x)-x{\gC}(x)^2)-xy({\gC}(xy)-xy{\gC}(xy)^2)} \nonumber \\
% &=\frac{ x{\gC}(x)-xy{\gC}(xy) }{ x-xy } & . \text{(using \eqref{eq-Cquad})}
% \end{align}
This allows us to bring the generating function for the summed lengths into the simple form
\begin{equation}
    \vL^{(p)}_n = [x^n y^p]\left\{ \frac{{\gC}(x)-y{\gC}(xy)}{ 1-y } \left[ \frac{{\gC}(x)-y{\gC}(xy)}{ 1-y } -1\right] \right\} ~.
\end{equation}
By applying the identity \eqref{eq-genfunargmult} we can begin to evaluate the coefficient of $x^n$ in this generating function,
\begin{align}
    \vL^{(p)}_n &= [x^n y^p]\left\{ \frac{1}{(1-y)^2}\left[ {\gC}(x)^2 +y^2{\gC}(xy)^2 -  2y{\gC}(x){\gC}(xy)\right] - \frac{1}{1-y} \left[ {\gC}(x) -y {\gC}(xy) \right] \right\} \nonumber \\
                &= [y^p]\left\{ \frac{1}{(1-y)^2}\left[ [x^n]\{{\gC}(x)^2\} +y^2 y^n [x^n]\{{\gC}(x)^2\} - 2y [x^n]\{{\gC}(x){\gC}(xy)\} \right] \right. \nonumber \\
                & \qquad \qquad \left. - \frac{1}{1-y} \left[ [x^n]\{{\gC}(x)\} -y y^n[x^n]\{{\gC}(x)\} \right] \right\} \nonumber\\
                &= [y^p] \left\{ \frac{1+y^{n+2}}{(1-y)^2}\vC_{n+1} - \frac{1+y^{n+1}}{1-y} \vC_n - \frac{2y}{(1-y)^2}  [x^n]\{{\gC}(x){\gC}(xy)\} \right\} ~,
\end{align}
the final term here can be evaluated using \eqref{eq-genfunprod}. 
We also notice here that $p$ is always less than $n$ (there is no way a path can contain more vertices than the tree it runs through) so we can omit the terms containing $y^n$. 
We recognise that $[y^p]\{(1-y)^{-2}\}=(p+1)$, therefore
\begin{align}
    \vL^{(p)}_n &= (p+1)\vC_{n+1} - \vC_n - 2 [y^p]\left\{ \frac{y}{(1-y)^2} \sum_{k=0}^n \vC_{n-k} y^k \vC_k\right\} \nonumber \\
                &= (p+1)\vC_{n+1} - \vC_n - 2 [y^p]\left\{ \sum_{k=0}^n \frac{y^{k+1}}{(1-y)^2} \vC_k\vC_{n-k} \right\} ~.
\end{align}
The sum in the final term here contains powers of $y$ running up to $n+1$. 
Restricting to terms with powers of y $\le p$, the summation becomes
\begin{align}
    \vL^{(p)}_n  &= (p+1)\vC_{n+1} - \vC_n - 2 \sum_{k=0}^{p-1}  [y^p] \left\{ \frac{y^{k+1}}{(1-y)^2} \right\}  \vC_k\vC_{n-k} \nonumber \\
                 &= (p+1)\vC_{n+1} - \vC_n - 2 \sum_{k=0}^{p-1}  [y^{p-k-1}] \left\{ \frac{1}{(1-y)^2} \right\}  \vC_k\vC_{n-k} ~.
\end{align}
We arrive at an expression for the summed length of paths from the root to the $p^\text{th}$ leaf in a binary tree with $n$ vertices.
\begin{equation}
    \vL^{(p)}_n  = (p+1)\vC_{n+1} - \vC_n - 2  \sum_{k=0}^{p-1}  (p-k)  \vC_k\vC_{n-k}  ~.
    \label{eq-leaflenghtsum}
\end{equation}

In order to reach a closed formula we need to remove the final sum, which is similar to Segner's relation
$%\begin{equation}
\sum_{k=0}^{n}\vC_k\vC_{n-k} = \vC_{n+1} %\quad,
$, % \end{equation}
but with an incomplete sum, i.e.
\begin{equation}
    \aS_{pn} = \sum_{k=0}^{p}\vC_{k}\vC_{n-k} ~.
\end{equation}
We claim that
\begin{equation}
    \aS_{pn} = \frac{1}{2}\vC_{n+1} + \frac{1}{2} \frac{ (2 p + 1 - n) (p + 2) (n - p + 1)}{(n + 1) (n + 2)} \vC_{p+1} \vC_{n-p} ~,
    \label{eq-incompSegner}
\end{equation}
which may be proven by induction, using only the recurrence relation for the Catalan numbers given in \eqref{eq-Catalan_recursion}.
%\begin{equation}
%    \vC_{n+1} = 2\frac{2n+1}{n+2}\vC_n \quad.
%\end{equation}
First we confirm validity for $p=0$, i.e.\
\begin{equation}
    \aS_{0n} = \frac{1}{2}\vC_{n+1} + \frac{1}{2} \frac{ (1 - n) (2) (n +1)}{(n + 1) (n + 2)} \vC_{1} \vC_{n}
             = \frac{2n+1}{n+2} \vC_{n} + \frac{ (1 - n) }{(n + 2)}\vC_n 
             = \vC_n = \sum_{k=0}^{0}\vC_{k}\vC_{n-k} ~.
\end{equation}
Assuming the expression to be valid for $p$, we proceed to show that this implies the validity for $p+1$, i.e.\
\begin{align}
    \aS_{p+1,n} &= \aS_{pn} + \vC_{p+1} \vC_{n-(p+1)} \nonumber \\
                &= \frac{1}{2} \vC_{n+1} + \left[ \frac{1}{2} \frac{(2p+1-n)(p+2)(n-p+1)}{(n+1)(n+2)} \vC_{n-p} + \vC_{n-p-1} \right] \vC_{p+1} \nonumber \\
%                &= \frac{1}{2} \vC_{n+1} + \left[ \frac{1}{2} \frac{(2p+1-n)(p+2)(n-p+1)}{(n+1)(n+2)} 2 \frac{2(n-p-1)+1}{(n-p-1)+2}\vC_{n-p-1} \vC_{n-p-1} \right] \vC_{p+1} \nonumber \\
                &= \frac{1}{2}\vC_{n+1} + \left[ \frac{(2p+1-n)(p+2)(2(n-p)-1)}{(n+1)(n+2)} \vC_{n-p-1} + \vC_{n-p-1} \right] \vC_{p+1} \nonumber \\
                &= \frac{1}{2}\vC_{n+1} + \frac{(2p+3-n)(n-p)(2p+3)}{(n+1)(n+2)} \vC_{n-p-1} \vC_{p+1} \nonumber \\
                &= \frac{1}{2}\vC_{n+1} + \frac{(2p+3-n)(n-p)(2p+3)}{(n+1)(n+2)} \vC_{n-p-1} \frac{1}{2} \frac{p+3}{2p+3} \vC_{p+1} \nonumber \\
                &= \frac{1}{2}\vC_{n+1} + \frac{1}{2} \frac{[2(p+1)+1-n][(p+1)+2][n-(p+1)+1]}{(n+1)(n+2)} \vC_{n-(p+1)} \vC_{p+1}
\end{align}
and so the formula is proven by induction.
As a final check we can look at the complete sum $\aS_{nn}$ which should reproduce Segner's relation
\begin{equation}
    \aS_{nn} = \frac{1}{2}\vC_{n+1} + \frac{1}{2} \frac{ (2 n + 1 - n) (n + 2) (n - n + 1)}{(n + 1) (n + 2)} \vC_{n+1} \vC_{n-n}
             = \frac{1}{2}\vC_{n+1}+\frac{1}{2}\vC_{n+1}\vC_{1}
             = \vC_{n+1} ~.
\end{equation}
Similarly, the relation
\begin{align}
% \aT_{pn} &=
     \sum_{k=0}^{p-1} (p-k) \vC_{n-k} \vC_{k}
     &= \frac{1}{(n+1)(n+2)} \left\{ (2n+1)(p+1)(n+1) \vC_{n} \right. \nonumber \\
     & \qquad \left. - (2 p+1)(p+1) \vC_{p} [2(n-p)+1] (n-p+1)\vC_{n-p}\right\} 
%    &= \frac{1}{(n+1)(n+2)} \left[ (p+1) \aR_{n} - \aR_{p} \aR_{n-p} \right]
    \label{eq-incompRautu}
\end{align}
%with $\aR_{n} = (2n+1)(n+1) \vC_{n}$.
can be proven by induction.
Finally, using \eqref{eq-incompRautu}, we find that \eqref{eq-leaflenghtsum} in explicit form gives the depth function \eqref{eq-av_depth}.
%\begin{align}
%    \vL^{(p)}_n  &= (p+1)\vC_{n+1} - \vC_n - \frac{2(2n+1)(p+1) \vC_{n}}{n+2} \nonumber \\
%                 &~+ \frac{ (p+1) (2p+1) (n-p+1) (2n-2p+1) \vC_{p} \vC_{n-p} }{(n+1) (n+2)} \nonumber \\
%                 &\!\!\!\!= \frac{2(p+1)(2p+1)(n-p+1)(2n-2p+1)}{(n+1)(n+2)} \vC_{p} \vC_{n-p} \nonumber \\
%                 & \qquad \qquad - \vC_{n}.
%\end{align}

%%%%%%%%%%%%%%%%%%%%%%%%%%%%%%%%%%%%%%%%%%%%%%%%%%%%%%%%%%%%%%%%%%%
\section*{References}
%%%%%%%%%%%%%%%%%%%%%%%%%%%%%%%%%%%%%%%%%%%%%%%%%%%%%%%%%%%%%%%%%%%
%\bibliographystyle{nature}
%\bibliography{bibliography/bibliograph}

\end{document}